\documentclass{ws-procs9x6}
\usepackage{latexsym}
\usepackage{relsize}
\usepackage{color}
\begin{document} 

\title{
UNIVERSALITY OF PHASE TRANSITION DYNAMICS:  TOPOLOGICAL DEFECTS FROM SYMMETRY BREAKING}
\author{ADOLFO DEL CAMPO$^{1,2,*}$ and WOJCIECH H. ZUREK$^1$}

\address{$^1$Theoretical Division,  Los Alamos National Laboratory, Los Alamos, NM 87545, USA}
\address{$^2$Center for Nonlinear Studies,  Los Alamos National Laboratory, Los Alamos, NM 87545, USA\\
}

\begin{center}
{\it To Tom W. B. Kibble, on occasion of his 80$^{th}$ birthday.}
\end{center}
%

\def\sgn{{\rm sgn}}
\def\H{{\rm H}}
\def\x{{\rm\bf x}}
\def\y{{\rm\bf y}}
\def\p{{\rm\bf p}}
\def\q{{\rm\bf q}}
\def\k{{\rm\bf k}}
\def\ii{{\rm i}}
\def\ee{{\rm e}}
\def\d{{\rm d}}
\def\la{\langle}
\def\ra{\rangle}
\def\om{\omega}
\def\Om{\Omega}
\def\vep{\varepsilon}
\def\wh{\widehat}
\def\tr{\rm{Tr}}
\def\da{\dagger}
\newcommand{\beqa}{\begin{eqnarray}}
\newcommand{\eeqa}{\end{eqnarray}}
\newcommand{\intf}{\int_{-\infty}^\infty}
\newcommand{\into}{\int_0^\infty}
\newcommand{\nut}{\nu_t}
\newcommand{\nutc}{ \nu_{tc}}

\begin{abstract}
In the course of a non-equilibrium continuous phase transition, the dynamics ceases to be adiabatic in the vicinity of the critical point as a result of the critical slowing down (the divergence of the relaxation time in the neighborhood of the critical point). This enforces a local choice of the broken symmetry and can lead to the formation of topological defects. The Kibble-Zurek mechanism (KZM) was developed to describe the associated nonequilibrium dynamics and to estimate the density of defects as a function of the quench rate through the transition.
During recent years, several new experiments investigating formation of defects in phase transitions induced by a quench both in classical and quantum mechanical systems were carried out.
At the same time, some established results were called into question.
We review and analyze the Kibble-Zurek mechanism focusing in particular on this surge of activity, and suggest possible directions for further progress. 
\end{abstract}


\bodymatter
%
%
%


\section{Introduction}


The aim of this paper is to provide a limited review of the experiments that test  the Kibble-Zurek mechanism (KZM): we shall focus on the experiments that test the scaling of the number of topological defects with the quench rate predicted by the KZM. This self-imposed restriction limits the number of the relevant experiments to a manageable  total. It is also a sign that the field -- that has its roots in the seminal papers of Tom Kibble \cite{Kibble76,Kibble80} -- has matured, so that the question that was initially most pressing (i.e., whether topological defects form at all via KZM) has been by now answered in the affirmative in a variety of systems 
\cite{Chuang91,Bowick94,Ruutu96,Bauerle96,Carmi00,Maniv03,Sadler06,Anderson08,Golubchik10,Schaetz13,EH13,Ulm13,
Tanja13,Lamporesi13}, although $^4$He remains a confounding exception 
\cite{Hendry94,Dodd98}.

The scaling of the defect density with the quench rate -- prediction of the {\it non-equilibrium} effect using {\it equilibrium} critical exponents\cite{Zurek85,Zurek93} -- is the key testable consequence of the KZM. However, the resulting dependence of the size of the domains where symmetry can be broken ``in unison'' is usually  given by a power law with a small fractional exponent. Therefore, to detect a significant variation in the defect density one needs to vary quench rates over a large range. This tends to be difficult in the traditional thermodynamic phase transition experiments. For instance, cooling (that can lead to a symmetry breaking transition) will typically result in temperature gradients inside the bulk of the system that can suppress defect formation\cite{KV97,DLZ99}, but it can also drive convection that can create defects, such as vortex lines in superfluids, independently of the KZM\cite{Hendry94,Dodd98}.

There are several reviews of the subject starting with \cite{Zurek96} and more recent monographs\cite{BZDA00,Kibble03,Kibble07,Dziarmaga10,Polkovnikov11}
that discuss the KZM, its consequences, and related phenomena in phase transitions. As is also the case with this review, all of these reviews cover only selected fragments of the field either because (as a result of recent developments) they are out of date, or because they are focused on specific subfields (e.g., quantum phase transitions). We focus on the (mostly recent) experiments that test scalings predicted by the KZM and the related theoretical developments.


\section{The Kibble-Zurek mechanism}


Consider the dynamics of spontaneous symmetry breaking
in the course of a phase transition induced by the change of a control parameter $\lambda$.
A continuous second-order phase transition is characterized by the divergence (usually as a power-law) of  both the {\it equilibrium} correlation length $\xi$
\beqa
\label{xieq}
\xi(\varepsilon)  =  \frac{\xi_{0}}{|\varepsilon|^{\nu}},
\eeqa
and  {\it equilibrium} relaxation time $\tau$ 
\beqa
\label{taueq}
\tau(\varepsilon)  =  \frac{\tau_{0}}{|\varepsilon|^{z\nu}},
\eeqa
as a function of the distance to the critical point  $\lambda_c$. It is convenient to  
define the reduced distance parameter 
\beqa
\varepsilon=\frac{\lambda_c-\lambda}{\lambda_c},
\eeqa
in terms of which the system initially prepared in the high-symmetry phase ($\varepsilon<0$) is forced to face a spontaneous symmetry breaking scenario 
as the critical point is crossed towards the degenerate vacuum manifold ($\varepsilon>0$).

In Eq. (\ref{xieq}), $\nu$ is the correlation length critical exponent, while $z$ in Eq.(\ref{taueq}) is the dynamic critical exponent.
Different systems belonging to the same universality class share the same critical exponents.
Above, $\xi_0$ and $\tau_0$ are dimensionful constants that depend on the microphysics in contrast with $\nu$ and $z$ that depend only on the universality class of the transition.
The Kibble-Zurek mechanism (KZM) describes the dynamics of a continuous phase transition under a time-dependent change of $\lambda$ across the critical value. 
The time-dependence $\lambda(t)$ in the proximity of $\lambda_c$ can usually be linearized.
Therefore, we assume a linear quench 
\beqa
\lambda(t)=\lambda_c[1-\varepsilon(t)]
\eeqa 
symmetric around the critical point so that the reduced parameter  is characterized by the quench time $\tau_Q$ and varies linearly in time according to 
\beqa
\label{varepst}
\varepsilon(t)=\frac{t}{\tau_Q},
\eeqa
in  $t\in[-\tau_Q,\tau_Q]$, the critical point being reached at $t=0$. 
\begin{figure}
\begin{center}
\psfig{file=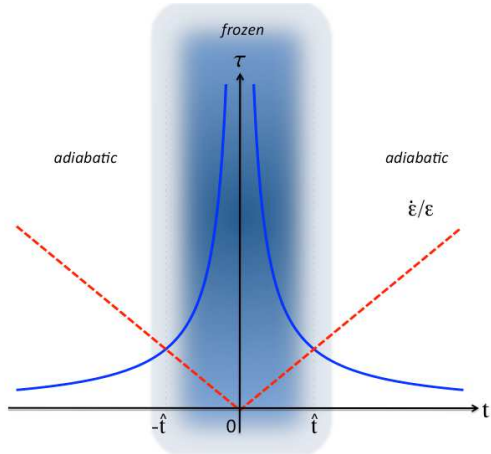,width=0.7\linewidth}
\end{center}
\caption{
Schematic representation of the freeze-out captured by the  adiabatic-impulse approximation. During a linear quench, the reduced control parameter $\varepsilon=t/\tau_Q$  forces the system to cross the critical point from the high symmetry phase ($t<0$) to the low symmetry phase ($t>0$). Due to divergence of the equilibrium relaxation time, associated with  the critical slowing down in the neighbourhood of $\varepsilon=0$, the order parameter of the system ceases to follow the equilibrium expectation value  and enters an impulse stage within the time interval $[-\hat{t},\hat{t}]$. 
}
\label{fig1}
\end{figure}
Far away from the critical point $|\lambda|\gg\lambda_c$, the equilibrium relaxation time is very small with respect to the time remaining until reaching the critical point  following the quench (\ref{varepst}), and the dynamics is essentially adiabatic. In the opposite limit, in the close neighbourhood of $\varepsilon(t)=0$, the dynamics is approximately frozen due to the divergence of the equilibrium relaxation time (critical slowing down). The system is then unable to adjust to the externally imposed change of the reduced control parameter $\varepsilon(t)$.
Exploiting this intuition\cite{Zurek85}, the KZM splits the dynamics into the sequence of three stages where the dynamics is adiabatic, effectively frozen, and adiabatic again, as $\varepsilon(t)$ is varied from $\varepsilon(t)<0$ to $\varepsilon(t)>0$. See figure \ref{fig1} for a schematic representation. 

This simplification, often referred to as the adiabatic-impulse approximation, 
captures the essence of the non-equilibrium dynamics involved in the crossing of the phase transition at a finite rate. The inability of the collective degree of freedom that defines the order parameter to keep up with the change imposed from the outside is the essence of the freeze-out. This does not mean that all of the evolution in the system stops, or even that the evolution of the order parameter ceases completely: the microstate of the system will of course evolve as dictated by its (time-dependent) Hamiltonian, and even the local thermodynamic equilibrium of the microscopic degrees of freedom may be maintained. However, the order parameter will cease to follow its equilibrium value, and it will be able to catch up with it locally, to the extent allowed  by the presence of topological defects, only after the critical point has been passed, usually with a delay of about $\hat t$, as illustrated, for example, by numerical simulations of BEC formation\cite{DSZ12}.

The boundary between the adiabatic and frozen stages can be estimated by comparing the equilibrium relaxation time with the time elapsed after crossing the critical point
\beqa
\tau(t)\approx |\varepsilon/\dot{\varepsilon}|=t.
\eeqa
This equation \cite{Zurek85} yields the time scale
\beqa
\hat{t}\sim\left(\tau_{0}\tau_{Q}^{z\nu}\right)^{\frac{1}{1+z\nu}},
\label{eq:freeze_out_time}
\eeqa
known as the freeze-out time.
The degrees of freedom of the system relevant for the selection of broken symmetry cannot keep up with the externally imposed change of $\varepsilon$, and, consequently, the order parameter of the system lags behind its equilibrium value corresponding to the instantaneous value of $\varepsilon$ within the interval $\varepsilon \in[-\hat{\varepsilon},\hat{\varepsilon}]$, where
\beqa
\hat{\varepsilon}\equiv |\varepsilon(\hat{t})|\sim\left(\frac{\tau_{0}}{\tau_{Q}}\right)^{\frac{1}{1+z\nu}}.
\eeqa

Spontaneous symmetry breaking entails degeneracy of the ground state. 
In an extended system, causally disconnected regions will make independent choices of the vacuum in the new phase.   
A summary of the topological classification of the resulting defects using homotopy theory is presented in the Appendix \ref{secSSB}. 
The KZM sets the average size of these domains  by the value of the equilibrium correlation length at $\hat{\varepsilon}$ \cite{Zurek85},
\beqa
\hat{\xi}\equiv\xi[\hat{\varepsilon}]=\xi_{0}\left(\frac{\tau_{Q}}{\tau_{0}}\right)^{\frac{\nu}{1+z\nu}}.
\label{KZMlength}
\eeqa
This is the main prediction of the KZM. 

This simple form of a power law of $\hat t$ (and, consequently, of $\hat \xi$) arises only when the relaxation time of the system scales as a power law of $\varepsilon$. This need not always be the case. For example, in the Kosterlitz-Thouless phase transition universality class, of relevance to 2D Bose gases, the critical slowing down is described by a more complicated (exponential) dependence on $\varepsilon$.
A more complex dependence of $\hat t$ and $\hat \xi$ on $\tau_Q$ (rather than a simple power law) 
would be then predicted as a result\cite{DZ13}.

The above estimate of the $\hat \xi$ is often recast as an estimate for the resulting density of topological defects,
\beqa
\label{dhkzm}
n\sim
\frac {{\hat \xi}^{d}} {\hat \xi^{D}} =\frac 1 { {\xi_0}^{D-d}} \left(\frac{\tau_{0}}{\tau_{Q}}\right)^{(D-d)\frac{\nu}{1+z\nu}},
\eeqa
where $D$ and $d$ are the dimensions of the space and of the defects (e.g., $D=3$ and $d=1$ for vortex lines in a 3D superfluid). This order-of-magnitude prediction usually overestimates the real density of defects observed in numerics. A better estimate is  obtained by using a factor $f$, to multiply $\hat \xi$ in the above equations, where $f\approx 5-10$ depends on the specific model\cite{LagunaZ1,YZ,ABZ99,ABZ00,ions2,DSZ12}. Thus, while KZM provides an order-of-magnitude estimate of the density of defects, it does not provide a precise prediction of their number. However, if one were able to check the power law above, one could claim that the KZM holds and show that the non-equilibrium dynamics across the phase transition is also universal.
This requires the ability to measure the average number of excitations after driving the system at a given quench rate, and repeating this measurement for different quench rates.


\section{Landau-Zener crossing as a quantum example of the KZM}


Landau \cite{Landau32} and Zener \cite{Zener32} (see as well Stueckelberg \cite{Stueckelberg32} and  Majorana \cite{Majorana32}) provided an analytical description of the diabatic excitation probability in a two-level quantum mechanical system described by a Hamiltonian $\hat{\mathcal{H}}_0$ in which the energy gap between the two states varies linearly in time.
Using dimensionless units for all variables, 
\beqa
\label{HLZ}
\hat{\mathcal{H}}_0=
\begin{pmatrix}
t/\tau_Q & 1  \\
1 & -t/\tau_Q(t) 
\end{pmatrix} = \frac{t}{\tau_Q} \sigma^z +  \sigma^x,
\eeqa
where $ \sigma^{x,y,z}$ are the usual Pauli matrices, and for which the instantaneous eigenbasis reads:
\beqa
|\uparrow(t)\ra&=&\sin (\theta/2) |1\ra +\sin (\theta/2) |2\ra,\nonumber\\
|\downarrow(t)\ra&=& -\sin (\theta/2) |1 \ra+\cos (\theta/2) |2 \ra. \nonumber
\eeqa
The angle $\theta\in[0,\pi]$ obeys the relations
\beqa
\cos  \theta=\frac{\varepsilon }{\sqrt{1+\varepsilon^2}},
\quad \sin  \theta=\frac{1 }{\sqrt{1+\varepsilon^2}}, \nonumber
\eeqa
in terms of the reduced variable
\beqa
\varepsilon=\frac{t}{\tau_Q}.
\eeqa
The exact energy gap is $E_{\uparrow}(t)-E_{\downarrow}(t)=\sqrt{1+\varepsilon^2}$.
The Landau-Zener (LZ)  formula states that the excitation probability decays exponentially with the quench time
\beqa
P=e^{-\frac{\pi}{2}\tau_Q}.
\eeqa
Above, time is measured in units given by the inverse of the gap in Eq. (\ref{HLZ}) at its minimum.
This results has been extended to multi-state problems \cite{DO67,BE93,DO95,DO00}
as well as nonlinear modulations of $\varepsilon(t)$ \cite{Ostrovsky03,Sinitsyn13}. 

Damski has shown that the quantum dynamics across a Landau-Zener (LZ) transition  is accurately described by 
the adiabatic-impulse approximation, and ultimately, by the KZM \cite{Damski05}.
The freeze-out time scale can be estimated by matching the inverse of the energy gap with the time scale $|\varepsilon/\dot{\varepsilon}|$
\beqa
\frac{1}{\sqrt{1+(\hat{t}/\tau_Q)^2}}=\alpha\hat{t}
\eeqa
where $\alpha$ is a constant.
It follows that $\hat{\varepsilon}=\hat{t}/\tau_Q=\frac{1}{\sqrt{2}}\bigg[\sqrt{1+\left(\frac{1}{\alpha\tau_Q}\right)^2}-1\bigg]^{1/2}$.
One can then consider the case where the system is initialized at a time $t_i\ll-\hat{t}$ and evolved until a final time $t_f\gg\hat{t}$.
The impulse stage occurs in the interval $[-\hat{t},\hat{t}]$ and the excitation probability can then be approximated by
\beqa
P=|\la\uparrow(\hat{t})|\downarrow(-\hat{t})\ra|^2=\frac{\hat{\varepsilon}^2}{1+\hat{\varepsilon}^2}.
\eeqa
Using the estimate for $\hat{\varepsilon}$, one finds that $P=1-\alpha\tau_Q/2+(\alpha\tau_Q)^2/2+\dots$.  
The optimal value $\alpha=\pi/2$ can be extracted from the comparison with the exact solution of the LZ problem \cite{DZ06}. This result agrees with the LZ formula up to third order in $\tau_Q$.

Exploiting the adiabatic impulse approximation, one can consider as well asymmetric quenches, such as when $t_i=0$, for which
\beqa
P=|\la\uparrow(\hat{t})|\downarrow(0)\ra|^2=\frac{1}{2} -\frac{1}{2\sqrt{1+\hat{\varepsilon}^2}}.
\eeqa
Its expansion, $P=\frac{1}{2}-\frac{1}{2}\sqrt{\alpha\tau_Q}+\frac{1}{8}(\alpha\tau_Q)^{3/2}+\dots$, matches well the exact result for $\alpha\simeq \pi/4$ \cite{DZ06}.

\begin{figure}
\begin{center}
\psfig{file=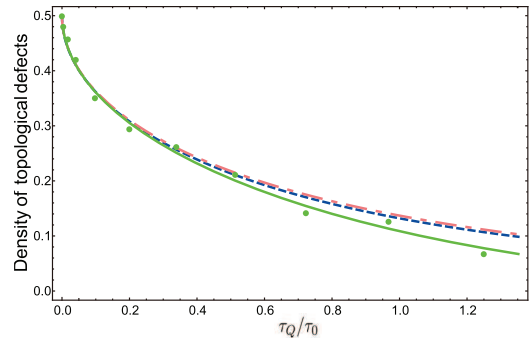,width=0.6\linewidth}
\end{center}
\caption{Experimental optical simulation of the quantum dynamics across a LZ crossing supporting the adiabatic-impulse approximation.   The  measured density of excitations (green dots) agrees with the exact solution (solid green line) and the estimate based on the adiabatic-impulse approximation\cite{Damski05}.
From {\it Xu et al.}\cite{KeyLab13}.}
\label{fig_MISF2}
\end{figure}
An experimental demonstration of the KZM-LZ connection \cite{Damski05}, the possibility of describing a LZ crossing using the adiabatic-impulse approximation which is a core feature of the KZM, has recently been achieved using a linear optical quantum simulator at the Key Laboratory of Quantum Information \cite{KeyLab13}. A second  experiment in the same center, 
this time in a semiconductor electron charge qubit, has further confirmed the universal validity of the adiabatic-impulse approximation \cite{KeyLab13b}.

\subsection{Controlling excitations in Landau-Zener crossing}

Excitations formed during a LZ crossing at an arbitrary finite-rate can be completely suppressed by counterdiabatic driving.
This method was introduced by Demirplak and Rice \cite{DR03},  and Berry \cite{Berry09}, and is also referred to as the transitionless quantum driving.
Provided that one can diagonalize the Hamiltonian of interest $\hat{\mathcal{H}}_0[\lambda(t)]$ (that is, find its instantaneous eigenstates $|n(\lambda)\ra$ and eigenvalues 
$E_n(\lambda)$) for every $\lambda(t)$, it is possible to enforce the dynamics exactly through the adiabatic manifold using counterdiabatic fields (i.e., the fields that allow one to cross the adiabatic-impulse regime fast, but without the usually inevitable excitations).
Indeed, the adiabatic approximation 
\beqa
\psi_n(t)= \exp\left(-\frac{i}{\hbar}\int_0^t E_n(t')dt'-\frac{1}{\hbar} \int_0^t \la n|\partial_{t'} n\ra dt'\right)|n(t)\ra
\eeqa
 to $\hat{\mathcal{H}}_0[\lambda(t)]$ becomes the exact solution of the time-dependent Schr\"odinger equation with the Hamiltonian  $\hat{\mathcal{H}}=\hat{\mathcal{H}}_0+\hat{\mathcal{H}}_1$,
 where  
\beqa
\label{H1full}
\hat{\mathcal{H}}_1&=&i \lambda'(t)  \sum_n[|\partial_\lambda n\ra\la n|-\la n|\partial_\lambda n\ra|n\ra\la n|].
\eeqa

Counterdiabatic driving has been demonstrated experimentally in an effective two-level system realized with a Bose-Einstein condensate in the presence of an optical lattice potential \cite{expCD1}. This type of assisted quantum adiabatic passage has also been implemented in an electron spin of a single nitrogen-vacancy center in diamond \cite{expCD2}.
For the LZ crossing with $\lambda(t)=t/\tau_Q$,  one finds that the counterdiabatic field reduces to 
\beqa
\hat{\mathcal{H}}_1=-\frac{1}{2\tau_Q}\frac{\Delta}{1+(t/\tau_Q)^2}\sigma^y. 
\label{H1LZ}
\eeqa

Counterdiabatic driving is currently finding an increasing number of applications in quantum control \cite{DR08}, quantum information processing \cite{expCD2}, BEC and ultra cold atom physics \cite{delcampo13}, and other fields \cite{Torrontegui13}.


\section{Quantum phase transitions}


We have seen that two-level systems constitute an ideal platform to test the adiabatic-impulse approximation, a key ingredient of the quantum KZM.
However, KZM also predicts the typical size of the domains in the broken symmetry phase resulting from a finite-rate crossing of a critical point, i.e., 
it estimates the average distance between topological defects. To analyze this aspect it is required to consider spatially extended systems.
A wide variety of condensed-matter systems and statistical mechanics models exhibiting quantum phase transitions offer a test-bed for these predictions.

Quantum phase transitions are characterized by abrupt changes in the ground-state properties of a many-body systems as a control parameter is tuned \cite{Sachdev}.
In experimental realizations this control parameter is typically an external field such as a magnetic field acting on spins, a laser field in trapped ion systems 
or and optical lattice potential in ultracold atom quantum simulators \cite{Sachdev, Lewenstein07}. 

The extension of the KZM to quantum phase transitions was elucidated by studying the dynamics in quasi-free fermion models \cite{ZDZ05,Dziarmaga05,Polkovnikov05,KCH11}, 
and is by now well-documented \cite{Dziarmaga10,Polkovnikov11}.
A paradigmatic example is the one-dimensional quantum Ising chain described by the Hamiltonian
\beqa
\hat{\mathcal{H}}=-\sum_{k=1}^N\left[g(t)\sigma_n^x+\sigma_n^z\sigma_{n+1}^z\right],
\eeqa
where $g(t)$ plays the role of a magnetic field, which has a critical point at $|g_c|=1$. 
Remarkably, this model describes certain magnetic condensed matter systems \cite{Isingcm} and its quantum emulation, e.g., in ion traps, is the subject of ongoing efforts \cite{Isingqs}.
A quantum phase transition occurs between a paramagnetic phase ($|g|>1$) and a doubly-degenerate ferromagnetic phase ($|g|<1$).

Consider the time-dependent quench $g(t)=-t/\tau_Q$ with $t\in(-\infty,0)$. One can quantify the breakdown of adiabaticity dictated by the KZM using the average number of excitations for a given quench rate ending at $g=0$,
\beqa
n=\frac{1}{2N}\sum_{k=1}^N[1-\la\sigma_n^z\sigma_{n+1}^z\ra].
\eeqa
 Using standard techniques (a combination of the Jordan-Wigner transformation and Fourier transform), Dziarmaga was able to rewrite the system as a set of independent Landau-Zener crossings \cite{Dziarmaga05}. In the  thermodynamic limit ($N\gg 1$), the density of kinks can then be approximated by
\beqa
n=\frac{1}{2\pi}\int_{-\pi}^{\pi} p_kdk,
\eeqa
where $p_k$ is the probability of excitation in each mode.
%
In view of the applicability of the adiabatic-impulse approximation to each level,  the dynamics across the critical point might be expected to be described by the KZM. 
The resulting amount of excitations is found to scale as $n\propto \tau_Q^{-1/2}$. This result is based on an exact solution of the dynamics for the Ising model \cite{Dziarmaga05,DZ06}. However, it can be extended to an arbitrary $D$ dimensional Hamiltonian $\hat{\mathcal{H}}[\lambda(t)]$,  with a quantum critical point characterized by critical exponents $\nu$ and $z$, leading to the estimate $n\propto \tau_Q^{-\frac{(D-d)\nu}{\nu z+1}}$, see \cite{Polkovnikov05} and the reviews \cite{Dziarmaga10, Polkovnikov11}. 


\section{Adiabatic crossing of quantum phase transition}


Counterdiabatic driving \cite{DR03,Berry09} has been extended to many-body systems and to quasi-free fermion models exhibiting a quantum phase transitions \cite{DRZ12}.
Consider the family of  $D$ dimensional model Hamiltonians,which can be decomposed into the sum of uncoupled $\k$-mode Hamiltonians,
\beqa
\hat{\mathcal{H}}_0 =\sum_{\k}\psi_{\k}^\dagger \left[ \vec{a}_\k (\lambda(t))  \cdot \vec{ \sigma}_\k \right] \psi_{\k}, 
\label{models}
\eeqa
where the Pauli matrices in the  mode $\k$ are $\vec \sigma_{\k} \equiv (\sigma_\k^x, \sigma_\k^y,\sigma_\k^z )$.  $\psi_{\k}^\dagger = (c_{\k,1}^\dagger,c_{\k,2}^\dagger)$ are fermionic operators. The function $\vec a_\k (\lambda) \equiv (a^x_\k (\lambda),a^y_\k (\lambda),a^z_\k (\lambda))$ is specific for each  model \cite{Dziarmaga10}. Examples of quantum critical models  within this family  are the Ising and XY models \cite{Sachdev} in $D=1$, and the Kitaev model in $D =1,2$  \cite{EK,1dKitaev}.   As quasi-free fermion models, they can be written down as a sum of independent Landau-Zener crossings.
The dynamics across the the quantum critical point  can be driven through the adiabatic solution associated with $\hat{\mathcal{H}}_0$ under the action of the modified Hamiltonian $\hat{\mathcal{H}}=\hat{\mathcal{H}}_0+\hat{\mathcal{H}}_1$, where the counterdiabatic term  is given by \cite{DRZ12}
\beqa
\hat{\mathcal{H}}_1 &=& \lambda'(t)\sum_{\k}\frac{1}{2  |\vec a_\k(\lambda)|^2} \psi_{\k}^{\dag} \left[ (\vec a_\k (\lambda) \times \partial_\lambda  \vec a_\k (\lambda)) \cdot \vec \sigma_{\k}.  \right] \psi_{\k}
\eeqa
The auxiliary Hamiltonian $\hat{\mathcal{H}}_1$ involves highly non-local pairwise interactions in the fermionic representation and  many-body interactions in the spin representation, accessible in quantum simulators  \cite{kbody,Barreiro11,Casanova12}. If the range of the auxiliary Hamiltonian $\hat{\mathcal{H}}_1$ is restricted to a value $M$ (which is equivalent to include up to $M$-body spin interactions), an efficient suppression of excitations occurs in modes with $k>1/M$, as explicitly verified in the 1D quantum Ising model \cite{DRZ12}.
Simpler forms of the auxiliary Hamiltonian $\hat{\mathcal{H}}_1$ are obtained whenever $\hat{\mathcal{H}}_0$ contains exclusively homogeneous spin interactions \cite{Takahasi13}, as in the Lipkin-Meshkov-Glick model \cite{LMG65}.


\section{The KZM and transitions between steady states} 

As we have noted already in the introduction, experimental tests of the power law scaling predicted by the KZM are difficult, since the exponent that governs the dependence of $\hat \xi$ on $\tau_Q$ is usually fractional, and often much less than 1. It is therefore no surprise that the earliest experiments that were devised to test the KZM scaling were carried out in transitions between distinct non-equilibrium steady states (rather than between different equilibria) in driven systems\cite{Ducci99, Casado01,Casado06,Casado07}, where implementing the quench is often easier. In such systems -- for example, in Rayleigh-B\'enard convection -- the broken symmetry can be associated with convective flows driven by thermal gradients in presence of an external potential (e.g., gravity). Topological defects are the imperfections in the arrangement of these far from equilibrium convective patterns. For example, the lattice of normally hexagonal B\'enard cells may exhibit lattice defects.

The effective field theory (such as a suitable version of the Ginzburg-Landau model) is often used to represent symmetry breaking associated with the formation of such steady-state structures. One can therefore expect (based on this Ginzburg-Landau connection) that some of the features of the dynamics of symmetry breaking predicted by the KZM for equilibrium phase transitions can be also detected in the transitions between distinct non-equilibrium steady states that exhibit different symmetries. This was indeed the case in the nonlinear optical system. \cite{Ducci99}. 
However, more recent experiments (see e. g. \cite{Miranda12}) present a richer and more complicated picture. Indeed, the very nature of such steady state phenomena (e.g., the fact that defects appear in an order parameter defined by the lattice of relatively large, B\'enard cell sized structures) suggests caution in applying the KZM to transitions between distinct non-equilibrium states that exhibit different broken symmetries. The concepts such as ``the relevant speed of sound'' or the ``sonic horizon'' and, especially, the ideas underlying renormalization group (that are natural in the equilibrium second-order phase transitions, where the KZM was developed) are not directly applicable to switching between distinct non-equilibrium steady states. 

This inapplicability of renormalization is not a concern in the thermodynamic or quantum phase transitions where many orders of magnitude usually separate, e.g., the healing length, from the microscopic scales that determine the basic physics. This scale separation allows for the independence of the physics that governs dynamics of the order parameter (and, hence, e.g., the size of the sonic horizon) from the underlying microphysics. However, when one cannot appeal to renormalization, scalings deduced from the KZM need not hold, or could be only an approximation. 

An interesting and instructive recent example of the extent to which the KZM can be used as a guide in such more general class of symmetry breaking phenomena even when the underlying dynamics does not yield itself to renormalization (or, indeed, to modeling of the order parameter in terms of partial differential equations) is offered by experiments \cite{Miranda13} and computer simulations \cite{Miranda13,Ashcroft13}. In this case what happens is in qualitative agreement with KZM, but does not follow its predictions in detail. A quantum example of an oversimplified model of the Bose-Einstein condensation that did seem to approximately follow the KZM even though the usual BEC order parameter did not enter the discussion, and the dynamics was represented by transitions between discrete -- as in Ref. \cite{Ashcroft13} -- states was also analyzed some time ago\cite{Anglin99}.  

In spite of these caveats, the transitions between steady states have provided suggestive  early evidence of KZM ``mean field'' scalings\cite{Ducci99}. Recent interesting work (see \cite{Miranda13,Ashcroft13}, and references therein) can be regarded as an attempt to formulate an extension of KZM that might be, possibly in only an approximate way, valid even where there is no scaling traceable to renormalization, and even where partial differential equations cannot be used to represent bifurcation-like processes under study.  


\section{Winding Numbers in Loops}\label{secLoop}

\begin{figure}
\begin{center}
\psfig{file=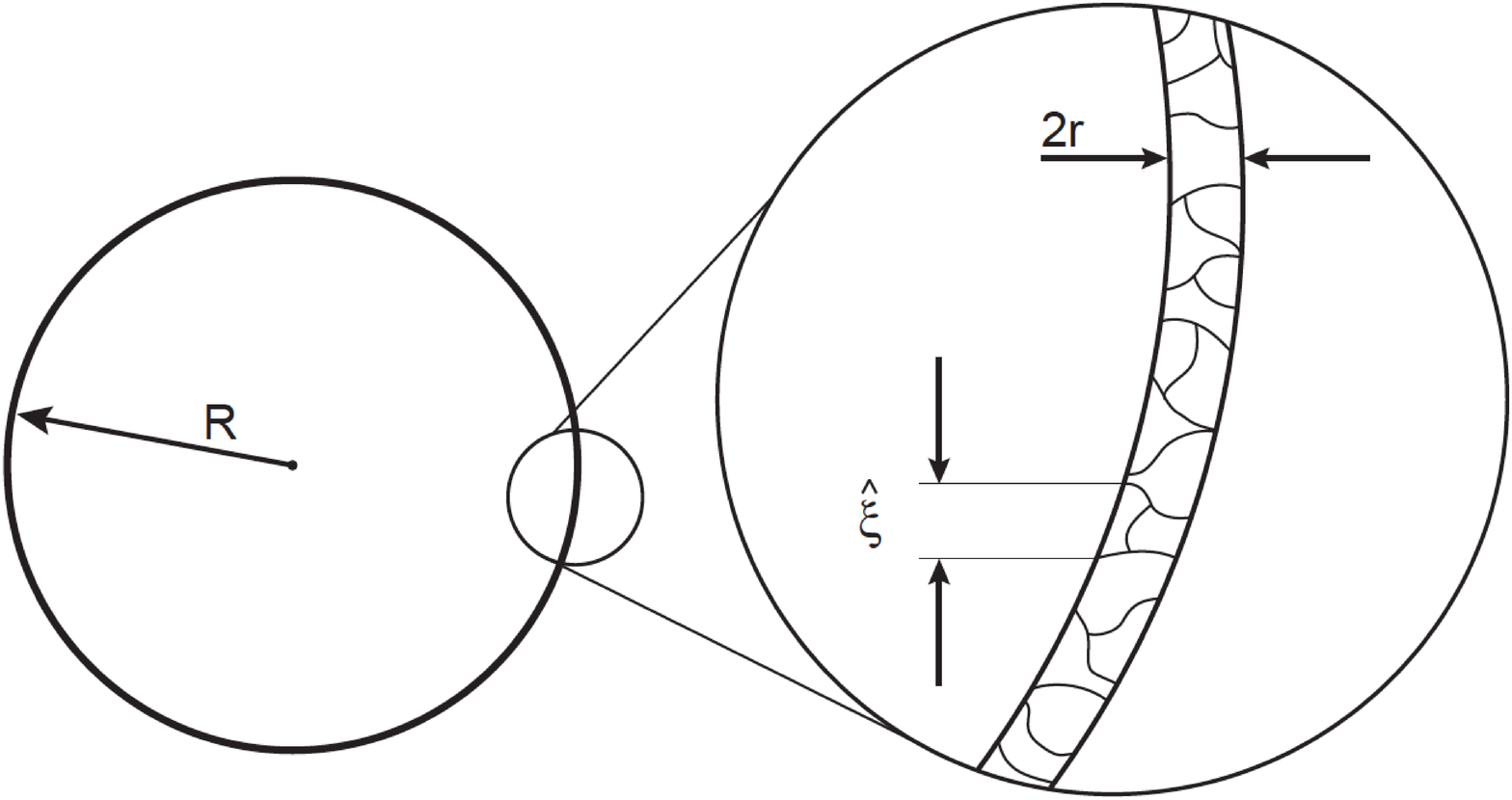,width=0.8\linewidth}
\end{center}
\caption{When a superfluid transition occurs in an annular container with the width comparable to the frozen-out healing length $\hat \xi$, distinct domains will choose the phase of the superfluid wavefunction independently. There will be then $N\simeq {\cal C} / {\hat \xi}$ of such domains with randomly chosen phase, or, in other words, along the circumference ${\cal C}=2\pi R$ a phase mismatch $\Delta \Theta \simeq \sqrt {{\cal C} / {\hat \xi}}$ will appear as a consequence of such random walk. The resulting phase gradient implies that a quantized, persistent flow can be induced by the KZM in a superfluid transition.}
\label{figSFring}
\end{figure}

The earliest prediction \cite{Zurek85} of scaling of the topologically nontrivial configurations induced by phase transition dynamics concerned winding numbers (and the resulting flows) in annular superfluid containers, see Fig. \ref{figSFring}.
The basic reasoning is straightforward: consider an annulus of circumference $\cal C$ that contains a substance which, as a result of a change in the external parameters, becomes a  superfluid (or superconductor). When the characteristic healing length set by the phase transition dynamics is $\hat \xi$, and ${\cal C} \gg \hat \xi$ while the width of the annulus is small so that it can be regarded as effectively one-dimensional loop, there will be
\beqa
N \simeq \frac {\cal C} {\hat \xi} 
\eeqa
sections of the annulus that independently select the phase of the condensate wavefunction. As a consequence of the resulting random walk of phase the typical net phase mismatch accumulated over the length ${\cal C}$ of the loop will be given by \cite{Zurek85}
\beqa
 \Delta \Theta \simeq \sqrt N \simeq \sqrt {\frac {\cal C} {\hat \xi}}.
 \label{33}
\eeqa
This net phase mismatch implies an average winding number:
\beqa
 {\cal W} \simeq \frac {\Delta \Theta} {2 \pi}.
 \label{34}
\eeqa 
After the phase ordering has smoothed out the domains, the resulting superfluid will flow with the velocity given by the phase gradient:
\beqa
  v = \left|\frac  \hbar m \vec \nabla \Theta\right| \simeq \frac  \hbar m \sqrt {\frac 1 { {\cal C} {\hat \xi}}}.
\eeqa
In the case of superconductors similar reasoning \cite{Zurek96} leads to magnetic field trapped inside $\cal C$ corresponding to the number of quanta given by $\cal W$. 


The basic idea of a random walk in phase resulting in the non-zero winding number has been successfully tested in the experiment involving a loop that was deliberately divided into $N=214$ superconducting sections by ``weak links'' \cite{Carmi00}. When the loop was reconnected into a single superconducting ring, flux quanta were trapped inside. Over many runs of this experiment, the resulting quantized magnetic flux had an approximately Gaussian distribution with the dispersion (related to the typical winding number) well approximated by the KZM-like Eqs. (\ref{33}) and (\ref{34}). 

By the very nature of the above reconnection experiment, the quench rate of the transition was irrelevant,  and, indeed, not well defined: the size of the ``domains'' that can choose the same phase was set by the distance between the weak links,  so that  the number of such domains was constant ($N=214$), and hence independent of the quench rate. 

The dependence of the typical trapped winding numbers on the quench rate is difficult to test in the laboratory. The expected power law is even smaller  (by a factor of 2) than the already small fractional power $\frac{\nu}{1+z\nu}$ that governs the size of $\hat \xi$. Moreover, the quench should be uniform -- it must happen nearly simultaneously in the whole annulus -- for, otherwise, the speed of the relevant sound may exceed the speed of the transition front, so the regions that ``go superfluid'' first will communicate their choice of the phase selection to the neighborhood, and the resulting winding numbers can be suppressed\cite{KV97,DLZ99}.

\begin{figure}
\begin{center}
\psfig{file=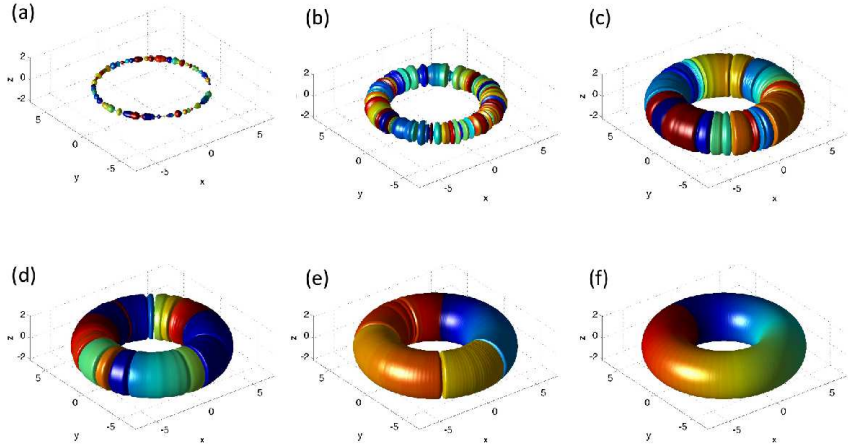,width=\linewidth}
\end{center}
\caption{Sequence of snapshots of isodensity surfaces during the growth of a BEC in a toroidal trap resulting in the formation of a superfluid current, modeled by the stochastic Gross-Pitaevskii equation  \cite{DSZ12}. 
The color coding describes the phase of the condensate along the ring. An early stage is characterized by large density and phase fluctuations. As the condensate growth there is a coarsening of both phase and  density fluctuations that result in the appearance of solitons. The final estate exhibits a uniform density and  winding number $\mathcal{W}=1$.}
\label{fig_torus}
\end{figure}

Numerical simulations of the stochastic Gross-Pitaevskii equation \cite{DSZ12},  such as those in figure \ref{fig_torus}, confirm this general paradigm and verify the KZM-predicted scalings. They also show how sensitive the resulting winding number is to the imperfections in the implementation of the transition. Such difficulties have so far hampered experimental verification of the KZM-predicted winding number scaling with the transition rate in, e.g., gaseous BEC's. 

\subsection{Trapping flux in small loops}

An interesting and successful set of increasingly sophisticated experiments that yielded a power law was carried out in small superconducting systems with the topology of an annulus: tunnel Josephson junctions and small superconducting loops \cite{Monaco02,Monaco03,Monaco06,Monaco08}. In this regime ${\cal C} \ll \hat \xi$, so that the winding numbers other than ${\cal W} = 0, \pm 1$ are exceedingly unlikely, and the natural observable in this case is the frequency of trapping a winding number $|{\cal W}| = 1$. 

As the random walk takes no more than one step, the square root of Eq. (\ref{33}) is no longer relevant, and it is reasonable to expect changes in the power law scaling with the quench rate. This general conclusion was reached using field-theoretic methods to predict doubling of the power law compared with the ${\cal C} \gg \hat \xi$ regime  \cite{KMR00}. Thus, when $\hat \xi \sim \tau_Q^{\frac 1 4}$ (as is expected in low-temperature superconductors), Ref.  \cite{KMR00} predicted exponent of $\frac 1 4$ for the scaling of typical winding numbers when  ${\cal C} \ll \hat \xi$ (as opposed to the exponent $\frac 1 8$ valid for   ${\cal C} \gg \hat \xi$, Eq. (\ref{33})).

\begin{figure}
\begin{center}
\psfig{file=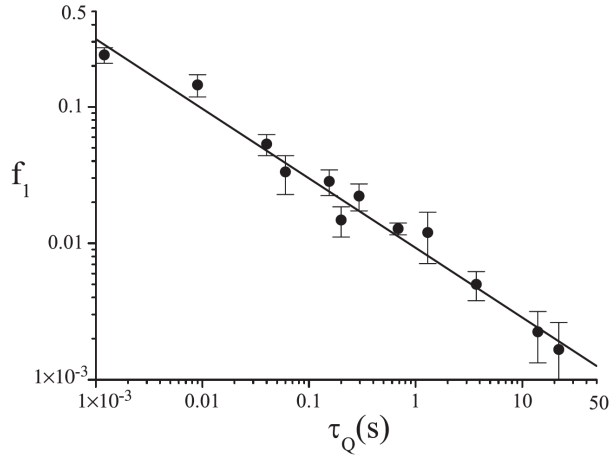,width=0.7\linewidth}
\end{center}
\caption{Scaling of the frequency $f_1$ of trapping a single fluxon with winding number $|\mathcal{W}|=1$ in annular Josephson tunnel junctions as a function of the quench rate $\tau_Q$. Each point is the result of averaging over many thermal cycles. A fit to a power law $f_1=a\tau_Q^{-\alpha}$ leads to $\alpha=0.51$. From {\it Monaco et al.} \cite{Monaco06}.  Copyright 2006 American Physical Society.}
\label{figFluxon}
\end{figure}
Initial experiments \cite{Monaco02} yielded  power law of the frequency of trapping a fluxon consistent with this prediction. However, later (and more refined and presumably more accurate) experiments \cite{Monaco06} resulted in a steeper slope with the exponent close to 0.5 as shown in figure \ref{figFluxon}, i.e. twice the prediction of Ref. \cite{KMR00}. This discrepancy was puzzling. Moreover, experiments on small superconducting loops by Monaco et al. \cite{Monaco09} reported similar scaling of the frequency of trapping with the exponent of $0.62 \pm 0.15$. This exponent is consistent with 0.5, again four times the slope expected for the scaling of typical $\cal W$ in the ${\cal C} \gg \hat \xi$ regime. The discrepancy with the initially anticipated scaling\cite{KMR00} in the tunnel Josephson junctions was attributed to the possible fabrication problems and the resulting ``proximity effect'' \cite{Monaco06}.

The resolution of the mystery that does not call on fabrication problems and resulting complications may be assisted by the recent observation \cite{Zurek13} that in the regime of small loops, ${\cal C} \ll \hat \xi$, dispersion of the winding numbers $\sqrt {\langle {\cal W}^2 \rangle }$ scales differently than ${\langle |{\cal W}| \rangle }$. Indeed, ${\langle |{\cal W}| \rangle }$ scales as probabilities (and, hence, frequencies) of $|{\cal W}| = 1$ while the dispersion scales as a square root of that probability. Therefore, it appears to us that the prediction of doubling of the scaling exponent of Ref. \cite{KMR00} is relevant to the dispersion $\sqrt {\langle {\cal W}^2 \rangle }$, while in the experiments that measure frequency of detection of $|{\cal W}| = 1$ one should expect four times the slope of the dispersion in the large-loop regime, ${\cal C} \gg \hat \xi$. With this revision\cite{Zurek13} of the original expectations\cite{KMR00}, the experiments on tunnel Josephson junctions as well as on the small superconducting loops are in excellent agreement with the predictions of the KZM and can be regarded as its verification (albeit in the mean field case).

The reason for the quadrupling (rather than just a simple doubling) of the power law for frequencies as well as for typical winding numbers characterized by ${\langle |{\cal W}| \rangle }$ is straightforward\cite{Zurek13}. We first note that the charges of topological defects created by the quench are anticorrelated\cite{LiuMazenko}. This is  reflected in the Eqs. (\ref{33}) and (\ref{34}) that recognize the phase of the condensate as the fundamental random variable. By contrast, if charges were assigned at random, typical $\cal W$ would be given by the square root of the number of defects subtended by the circumference $\cal C$. Thus, for ${\cal C} \gg \hat \xi$, when the contour contains many defects of both charges, random distribution of charges would be directly proportional to the circumference (rather than its square root, Eq. (\ref{33})).

\begin{figure}
\begin{center}
\includegraphics[width=0.8\linewidth]{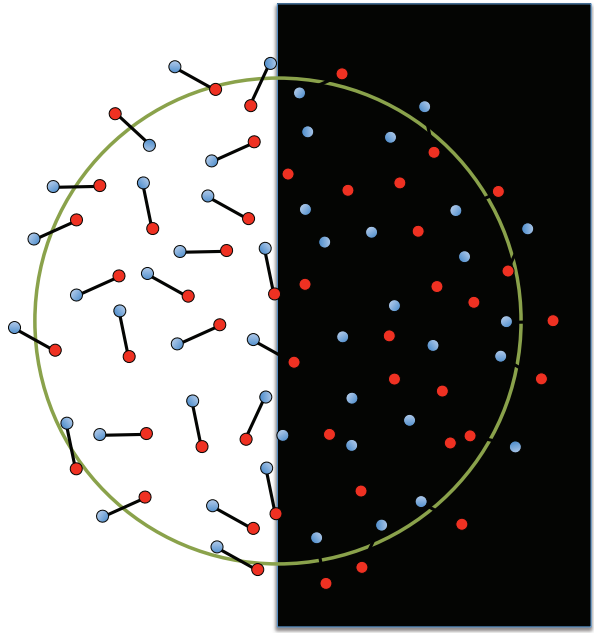}
\end{center}
\caption{\label{kzmpairs} The winding number for the circumference $\cal C$ (the circle above) due to vortex-antivortex pairs scattered randomly on a plane. Contribution of the pairs that are completely outside or completely inside $\cal C$ vanishes: only pairs that straddle the contour contribute to $\cal W$. The number of such pairs is proportional to the circumference $\cal C$. 
Note that pairing illustrated above is in a sense imaginary (as is suggested by the right hand side of the figure, where pair assignements are invisible), as there is generally no unique ``correct'' way to combine vortices and antivortices into pairs. Nevertheless, the recognition of pairing leads to correct scaling of winding numbers with $\cal C$.
When loops are so small that typically, at most only one end of a pair ``fits inside'', scaling changes, see Fig. \ref{tilt}.
From Zurek \cite{Zurek 13}. \copyright IOP Publishing. Reproduced by permission of IOP Publishing. All rights reserved.
}
\end{figure}

This scaling of typical $\cal W$ with $\cal C$ can be recovered in a simple model\cite{Zurek13} where pairs of the oppositely charged defects are randomly scattered on a plane (see Fig. \ref{kzmpairs}). 
The sizes of the pairs, as well as their separations, are presumably of the order of $\hat \xi$. The typical winding number is then given by the square root of the number of pairs disected by the $\cal C$, which leads to the scaling of Eq. (\ref{33}). 

\begin{figure}
\begin{center}
\includegraphics[width=0.8 \linewidth]{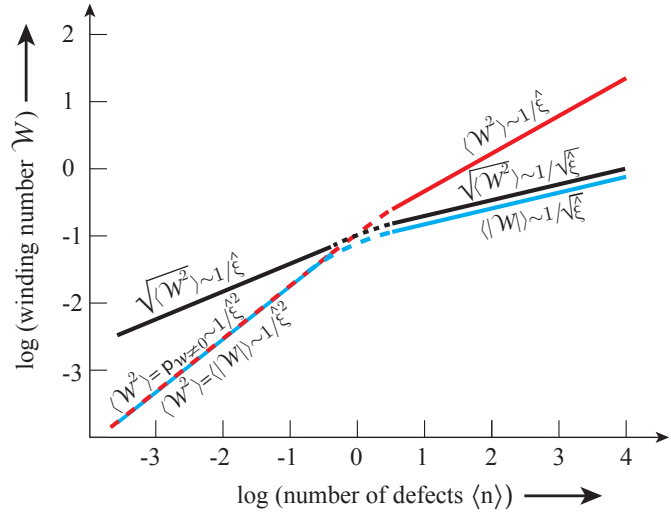}
\end{center}
\caption{\label{tilt} 
The tilt of the scaling of the dispersion, $\sqrt {\langle {\cal W}^2 \rangle }$, its square, and the average $\langle | {\cal W} | \rangle$ of the winding numbers is expected to change with the number of defects trapped inside $\cal C$ when $\langle n \rangle \simeq 1$\cite{Zurek13}. For loops that trap many defects, $\langle n \rangle \gg 1$, the dispersion and the average absolute value of $\cal W$ scale similarly, 
$\sqrt {\langle {\cal W}^2 \rangle} \sim | {\cal W} | \sim \sqrt {{\cal C}  /{\hat \xi} }$. 
However, different tilts, corresponding to the exponents that control the slopes of the dispersion and $\langle | {\cal W} | \rangle$, set in as $\langle n \rangle \ll 1$. Compared to $ \sqrt {{\cal C}  /{\hat \xi} }$,  the slope of the dispersion {\it doubles},  $\sqrt{\langle {\cal W}^2 \rangle }  \sim {\sqrt A_{\cal C} / {\hat \xi}}$ while the slope of the average absolute value {\it quadruples} so that $\langle | {\cal W}| \rangle  \simeq p_{|W| = 1} \sim A_{\cal C} / {\hat \xi}^2 \sim {\langle {\cal W}^2 \rangle }  $ when $\langle n \rangle \ll 1$, where $A_{\cal C}$ is the area enclosed by the contour ${\cal C}$. (Note that $\langle n \rangle \approx A_{\cal C} / {\hat \xi}^2$). From Zurek \cite{Zurek 13}. \copyright IOP Publishing. Reproduced by permission of IOP Publishing. All rights reserved.
}
\end{figure}

The heuristic picture of the generation of the winding number is then straightforward. The quench results in a random configuration of the order parameter deposited inside $\cal C$. Instead, we can imagine an infinite plane with all configurations of the order parameter left behind by the transition pockmarked by defects and sampled at random by dropping the contour $\cal C$ at random locations. When defects are paired up (as their anticorrelations suggest) and ${\cal C} \gg \hat \xi$, the scaling of Eq. (\ref{33}) is easily recovered (see Fig. \ref{tilt}). 
By contrast, when ${\cal C} \ll \hat \xi$, most of the loops tossed on the plane will end up ``empty'', hence, will have ${\cal W}=0$. Only on rare occasions when the loop of area $A_{\cal C} \sim {\cal C}^2 \ll {\hat \xi}^2$ ``traps'' a defect inside, the winding number will be $+1$ or $-1$, depending on the defect charge. Moreover, the probability of trapping the defects will scale as $\sim A_{\cal C} \times {\hat \xi}^{-2}$, as the density of the KZM defects is $\sim 1/{\hat \xi}^2$. Consequently, the probability (and, hence, the frequency) of finding a loop with $| {\cal W}| = 1$ in the case of $\langle |{\cal W} | \rangle \ll 1$ 
scales as:
\beqa
p_{{\cal W} = \pm 1} \sim \frac {A_{\cal C}} {{\hat \xi}^{2}} \sim \frac {{\cal C}^2} {{\hat \xi}^{2}} \ .
\eeqa 
Note that the power with which $\hat \xi$ appears above is {\it four} times the power law relating typical $\cal W$ and $\hat \xi$ when ${\cal C} \gg \hat \xi$, Eq. (\ref{33}). It follows that the scaling of the frequency (or of $\langle |{\cal W}| \rangle$) with $\tau_Q$ in this ${\cal C} \ll \hat \xi$ regime is four times steeper than for large (${\cal C} \gg \hat \xi$) loops. Scaling of $\langle {\cal W}^2 \rangle$ in this regime is equally steep, as $\langle {\cal W}^2 \rangle \approx \langle |{\cal W}| \rangle \approx p_{{\cal W} = \pm 1} \simeq {\cal C}^2/{\hat \xi}^2$. On the other hand, scaling of $\sqrt{\langle {\cal W}^2 \rangle} $ will only double (which is what may have been predicted by Ref.\cite{KMR00}). It is that discrepancy between the scaling of dispersion and frequency of detection in case of small loops that may account for the experimental results seen in Fig. \ref{figFluxon}.

This quadrupling is a combination of two doublings (or, rather, it reverses the consequences of two square roots that appear as the size of the loop increases from ${\cal C} \ll \hat \xi$ to ${\cal C} \gg \hat \xi$). For small loops, the frequency of trapping a single defect is proportional to the area inside $\cal C$, and this yields a proportionality to the area for $\langle {\cal W}^2 \rangle \approx \langle |{\cal W}| \rangle \approx p_{{\cal W} = \pm 1} \simeq {\cal C}^2/{\hat \xi}^2$. By contrast, for large loops the net winding number is given by the random walk in the phase (which yields square root \#1) of the number of pairs intercepted by ${\cal C} \gg \hat \xi$ rather than the area inside, ${A_{\cal C}}$ (which implies square root \#2). This change of the power law that governs the scalings will be reflected in the power law dependence of the winding number on the quench time $\tau_Q$.


\section{Defect formation in multiferroics}

\begin{figure}
\begin{center}
\psfig{file=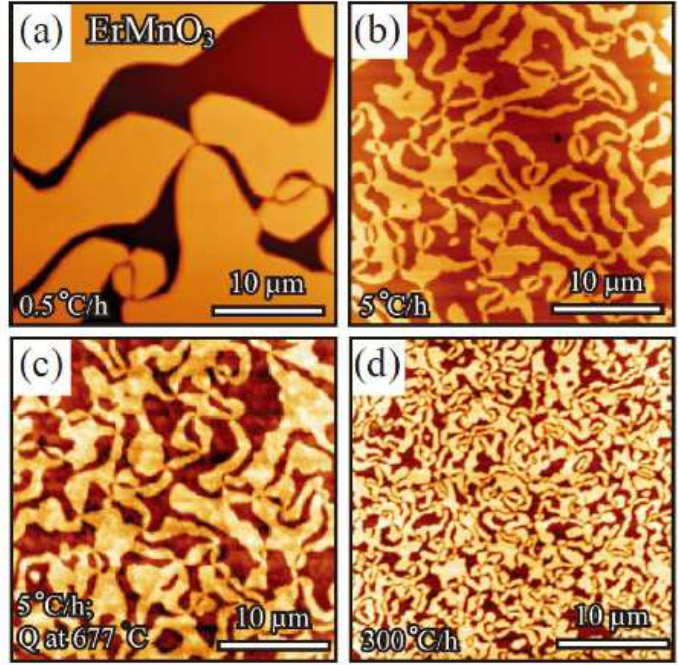,width=0.8\linewidth}
\end{center}
\caption{
Vortices (present where three dark and three bright domains merge) punctuate domain patterns formed in  chemically-etched ErMnO$_3$ crystals for different cooling rates. The characteristic scale extracted from the experiments (e.g., from the density of vortices) exhibits a scaling with the rate of quench that is consistent with the KZM predictions for the 3D XY model\cite{Griffin12}, which has $\nu = 0.6717$ and $\nu z = 1.3132$ calculated using Monte Carlo simulations \cite{3DXY}. 
From {\it Chae et al.} \cite{Chae12}. Copyright 2012 American Physical Society.}
\label{fig_mf}
\end{figure}

Multiferroics are materials that exhibit more than one primary ferroic order parameter simultaneously (i.e. in a single phase). Recent measurements in rare earth multiferroics have provided what may be a compelling evidence of the KZM\cite{Chae12}. The reason for excitement is illustrated in Fig. \ref{fig_mf}. It shows snapshots of the surface of ErMnO$_3$ cooled, at different rates, from about 1200$^\circ$C (i.e., from above the phase transition that occurs at ~1120-1140$^\circ$C) to room temperature. The mosaic pattern seen in this figure represents domains that form as a result of symmetry breaking. These domains are punctuated by vortex-like defects that appear where several domains meet. The topological charge of the point defect is determined by the order in which distinct phases are arranged. Clearly, the scale of the structures (that can be deduced from the density of the point defects) increases with the cooling time, as is shown in Fig. \ref{fig_mf}. 

The power-law exponent governing the dependence of the distance between the point defects in the quench rate is close to 0.25, which suggests a description in terms of the KZM with, e.g., the mean-field critical exponents $\nu = \frac 1 2$, $z = 2$ that would result in $\hat \xi $ scaling with the power $\frac \nu { 1 + \nu z}  = \frac 1 4$. However, Griffin et al.\cite{Griffin12} note that the universality class of the phase transition is the same as of the 3D XY model, which has $\nu = 0.6717$ and $\nu z = 1.3132$ calculated using Monte Carlo simulations \cite{3DXY} (as a caveat, note that the choice of $z$ depends on the dynamics, which is not well characterized in this case). Consequently, the predicted KZM exponent would be $\frac \nu { 1 + \nu z}  \simeq 0.29$. This is consistent with the experimentally measured value. Indeed, not just the slope of the power law dependence but also the net defect density are in approximate agreement with the {\it ab initio} calculations\cite{Griffin12}. 

While the above discussion can be interpreted as a resounding confirmation of the KZM, there are reasons for caution. To begin with, apart from the approximate critical temperature, very little is known about the actual critical behavior of ErMnO$_3$ (and similar rare earth manganites given by the chemical formula RMnO$_3$, where R stands for a rare earth element). This problem is largely due to the very high critical temperature, which makes, at least to date, the measurements that would allow one to extract $\nu$ and $z$ all but impossible. Indeed, at present it is not even completely clear, experimentally, whether the transition is of first or second order. 

The other reason for caution comes from the fast temperature quenches carried out recently\cite{Griffin12}. They have yielded (albeit in YMnO$_3$, and not in ErMnO$_3$ where the original study \cite{Chae12} was conducted) a surprise: 
the increase in the rate of much faster quenches (with cooling rates of up to $10^2$K/s) actually suppressed defect production, resulting in increasing size of domain sizes structures. This has not been yet explained, although several possibly relevant effects have been discussed \cite{Griffin12}. At present, it seems reasonable to wait for experimental confirmation of this `anti-KZM' effect before attempting to advance a detailed theory.

One might hope that the  experimental difficulties and the related uncertainties may be eventually overcome. Precision measurements of the scaling of $\hat \xi$ with $\tau_Q$ could be then increased sufficiently so that one might confirm that it is indeed close to 0.29 predicted by the 3DXY universality class (and not ``mean field''), and that would be a major coup. 


\section{The inhomogeneous Kibble-Zurek mechanism}\label{secIKZM}


Tests of the Kibble-Zurek mechanism in the laboratory  often face the situation in which the phase transition is inhomogeneous as opposed to being crossed everywhere at once \cite{DKZ13}. 
What survives from the Kibble-Zurek mechanism in inhomogeneous phase transitions is decided by causality\cite{Zurek09}. 
This realization has provided a foundation for an active area of research in recent years, with theory works\cite{DLZ99,ZD08,Zurek09,DZ09,DM10,DM10b,ions1,ions2,DRP11,WDGR11,Collura}  accompanied by a substantial experimental progress\cite{EH13,Ulm13,Tanja13,Lamporesi13} following the pioneering suggestion by Kibble and Volovik \cite{KV97}, who first focused on the problem of phase ordering behind a propagating front of a continuous phase transition.
This situation can arise as a result of a inhomogeneous tuning  of the control parameter driving the transition $\lambda=\lambda(x,t)$.
Alternatively, it might result from a spatial dependence of the critical value $\lambda_c=\lambda_c(x)$ which often occurs in trapped systems. Using the local density approximation, one could then replace the critical value $\lambda_c[\rho]$ for homogeneous density $\rho$ by $\lambda_c[\rho(x)]$. 
We refer the reader to the recent review for a detailed exposition of the subject \cite{DKZ13}.

\begin{figure}
\begin{center}
\psfig{file=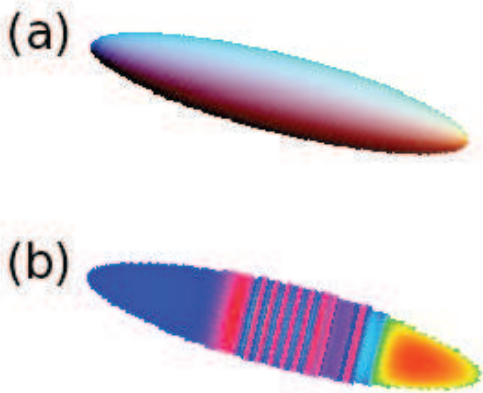,width=0.7\linewidth}
\end{center}
\caption{Bose-Einstein condensation by evaporative cooling in a harmonic trap offers an example of the inhomogeneous phase transition where causality enhances the dependence of the defect density on the quench rate. (a) Isodensity surface of Bose gas in a trap. Its density is highest in the center of the trap, and that is where the condensation will start when the cloud cools e.g. by evaporative cooling. When the region that becomes BEC first can communicate its choice of the condensate phase to the neighbouring domains, defects will not form. (b) On the other hand, when the speed of relevant sound $\hat s = \hat \xi / \hat t$ is less that the speed $v_F$ with which the critical point propagates as a result of cooling, different phases will be chosen by different domains (as indicated by the color coding) and grey solitons can be created.
From Zurek \cite{Zurek09}. Copyright 2009 American Physical Society.}
\label{Cigar}
\end{figure}

Let us assume that the critical point exhibits a spatial dependence  $\lambda_c=\lambda_c(x)$  and that the system undergoes a  homogeneous quench of the  control parameter with constant rate $1/\tau_Q$, 
\beqa
\lambda(t)=\lambda_c\left(1-\frac{t}{\tau_Q}\right),
\eeqa 
during the time interval  $t\in[-\tau_Q,\tau_Q]$. 
As in the homogeneous case, it is convenient to introduce the reduced control parameter 
\beqa
\label{rcp}
\varepsilon(x,t)=\frac{\lambda_c(x)-\lambda(t)}{\lambda_c(x)},
\eeqa
which takes values $\varepsilon(x,t)<0$ in the high symmetry phase where the system is initially prepared, it reaches $\varepsilon(x,t)=0$ at the critical point, and the broken-symmetry phase for $\varepsilon(x,t)>0$.

To establish the relationship with the homogeneous case, it is further convenient to introduce 
 an effective local quench time, 
\beqa
\tau_Q(x)=\left|\frac{\partial \varepsilon(x,t)}{\partial t}\right|^{-1}.
\eeqa
The condition  $\varepsilon(x_F,t_F)=0$ allows us to find the time $t_F$ at which the propagating front crosses the transition at the location $x_F=x$
\beqa
t_F(x)=\tau_Q\bigg[1-\frac{\lambda_c(x)}{\lambda_c(0)}\bigg],
\eeqa
in terms of which the reduced control parameter reads $\varepsilon(x,t)=\frac{t-t_F(x)}{\tau_Q(x)}$.
Matching, in the spirit of Eq. (\ref{taueq}), 
\beqa
\tau[\varepsilon(x,t)]=\bigg|\frac{\varepsilon(x,t)}{\dot{\varepsilon}(x,t)}\bigg|=|\varepsilon(x,t)|\tau_Q(x),
\eeqa
one obtains \cite{Zurek09,DRP11 }that
\beqa
\hat{\varepsilon}(x)=\bigg[\frac{\tau_0}{\tau_Q(x)}\bigg]^{\frac{1}{1+\nu z}}.
\eeqa
See \cite{DKZ13} for alternative derivations.
We note that $\hat{\varepsilon}(x)$ is associated with the local freeze-out time
$\hat{t}(x)=[\tau_0 \tau_Q(x)^{\nu z}]^{\frac{1}{1+\nu z}}$ measured with respect to $t_F(x)$ (that is, freeze-out will take place in the interval $[t_F(x)-\hat{t}(x),t_F(x)+\hat{t}(x)]$). 
It follows that the typical size of the domains in the broken symmetry phase is given by
\beqa
\hat{\xi}(x)\equiv\xi[\hat{\varepsilon}(x)]\simeq
\xi_0\bigg[\frac{\tau_Q(x)}{\tau_0}\bigg]^{\frac{\nu}{1+\nu z}}.
\eeqa
In contrast to the homogeneous scenario, defect formation does not occur all over the spatial extent $L$ of the system but it is restricted by causality\cite{Zurek09}.
Once a choice of a ground-state of the vacuum manifold is made locally in a given part of the system, it can be communicated to neighbouring regions. 
The characteristic local velocity of the perturbations, which determines the speed at which this choice can be communicated, is given by the analogue of the second-sound velocity in $^4$He that can be upper bounded by the ratio of the local frozen-out correlation length $\hat{\xi}(x)$
 and relaxation time scale $\hat{\tau}(x)=\tau[\hat{\varepsilon}]=\hat{t}(x)$,
this is,  by \cite{KV97} 
\beqa 
\label{hats}
\hat s =  \frac {\hat \xi} {\hat \tau} = \frac {\xi_0} {\tau_0} \bigg[\frac {\tau_0} {\tau_Q(x)} \bigg]^{\frac {\nu(z-1)} {1+\nu z}}.
\eeqa
When $\hat s$  is larger that the transition front velocity $v_F$, defect formation is suppressed.
The speed of propagation of this front can be estimated  to be \cite{Zurek09}
\beqa
v_F= \left|\frac {d x_F} {d t_F}\right| = \frac {\lambda_c(0)} {\tau_Q} \left| \frac {d \lambda_c(x)}{d x_F} \right|^{-1}=\left| \frac {d \tau_Q(x)}{d x_F} \right|^{-1}.
 \label{eq:v_F}
\eeqa
This expression diverges for homogeneous system, or where the system is locally homogeneous (e.g., whenever $\lambda_c(x)$ reaches an extremum).
For defects to be formed,  $\hat s<v_F$ is required. 

Numerical and analytical tests have confirmed this intuition, and thus, the role of causality in defect formation both in classical \cite{DLZ99,ions1,ions2,WDGR11}  
and quantum systems \cite{DM10,DM10b}.
This inequality is generally satisfied in a fraction of the system $\hat{X}=\hat{x}/L$, with $\hat{x}=\{x|v_F>\hat s\}$.
Within $\hat{x}$, the number of defects can be estimated using $\hat{\xi}(x)$.
The resulting density of defects in the whole system is then simply given by the total numbers of defects formed with the homogeneous density in regions where $v_F>\hat s$ divided by the total  system size, which in the 1D case reduces to 
\beqa
n\simeq \frac{1}{L}\int_{\{x|v_F>\hat s\}} \frac{1}{\xi_0}\bigg[\frac{\tau_0}{\tau_Q(x)}\bigg]^{\frac{\nu }{1+\nu z}}dx.
\label{dikzm}
\eeqa
This expression does not generally lead to a power-law in the quench rate \cite{DRP11}.
A power-law does however result in limiting cases \cite{Zurek09,DRP11}.
Whenever $\lambda_c(x)$ attains an extremum, say at $x=0$, it can be linearized as
\beqa
\lambda_c(x)=\lambda_c(0)+\frac{\lambda_c''(0)}{2}x^2+\mathcal{O}(x^3),
\eeqa
and the front velocity simplifies to
\beqa
\label{vfa}
v_F\simeq\frac{\lambda_c(0)}{\tau_Q |x\lambda_c''(0)|},
\eeqa
which diverges at the origin $x=0$.
The effective region of the system where defect formation is allowed by causality can be then estimated by comparing (\ref{hats}) and (\ref{vfa}).
Assuming $\hat x$ is simply connected and small enough, so that  $\tau_Q(x)\approx\tau_Q(0)$ within $\hat x$, it is found that
\beqa
|\hat{x}|\simeq\frac{\lambda_c(0)}{ |\lambda_c''(0)|\xi_0}\bigg[\frac {\tau_0} {\tau_Q(0)} \bigg]^{\frac {1+\nu} {1+\nu z}}, 
\eeqa
which increases with the quench rate, as expected.
This results in the total density of defects 
\beqa
n\simeq \frac{1}{L}\frac{\lambda_c(0)}{ |\lambda_c''(0)|\xi_0^2}\bigg[\frac {\tau_0} {\tau_Q(0)} \bigg]^{\frac {1+2\nu} {1+\nu z}},
\label{dikzm}
\eeqa
with a new power-law exponent\cite{Zurek09} which is a multiple, by a factor $\frac{1+2\nu}{\nu}$, of what is predicted for the density (e.g., $\hat{\xi}^{-1}$) by the homogenous KZM in 1D, Eq. (\ref{dhkzm}).
This constitutes a testable prediction of the inhomogeneous KZM (IKZM).
Numerical evidence supporting this scaling was first described in  \cite{ions1,ions2}, see as well \cite{WDGR11}.
A flurry of experimental activity testing the IKZM has been reported during 2012 and 2013 on the scaling of defect formation in inhomogeneous system and we now turn our attention to it.


\section{Kink formation in ion chains}
\label{secKink}

\begin{figure}
\begin{center}
\psfig{file=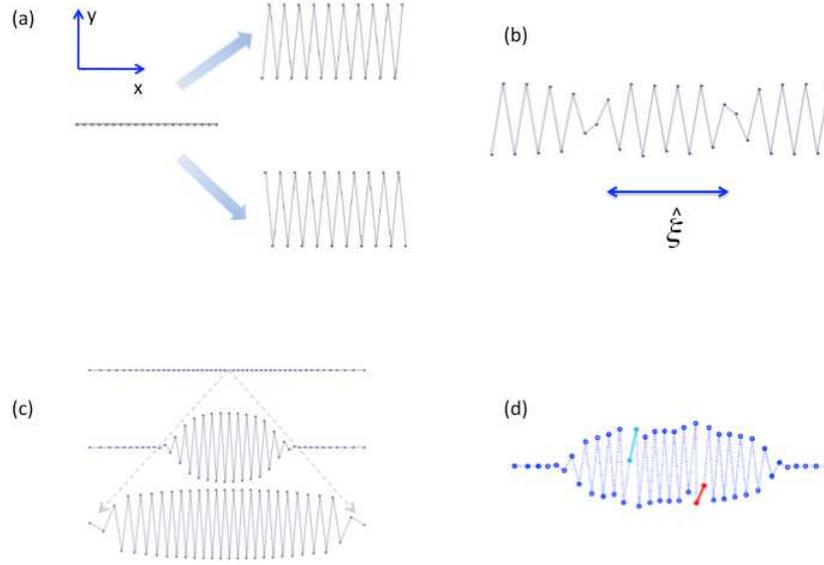,width=1\linewidth}
\end{center}
\caption{
Linear to zigzag phase transition in ion chains.
(a) Symmetry breaking in an homogeneous ion chain following a decompression of the transverse confinement. The vacuum manifold consists of two doubly-degenerate disconnected regions, associated with a $\mathbb{Z}_2$-symmetry breaking scenario. 
(b) Boundaries between disparate choices of the vacuum lead to the formation of $\mathbb{Z}_2$-kinks or domain walls.
(c) A sequence of ground state configurations in a harmonically trapped ion chain. As a result of the axial harmonic confinement the inter-ion spacing $a(x)$ is lowest at the center of the chain and increases sideways. Under a (homogeneous) decompression of the transverse confinement, the zigzag phase is first formed in the center of the chain (where Coulomb repulsion is higher) and coexists with region in the linear phase.
(d) The transverse decompression (or axial compression) of an inhomogeneous ion chain at finite rate can lead to the formation of structural defects. These defects are not stationary and can  propagate along the chain and annihilate by collisions (between a kink and an anti-kink with opposite topological charge) or can be lost at the edges of the chain.
}
\label{fig_ions}
\end{figure}

Coulomb crystals made of ion chains stand out as a platform for quantum technologies as a result of their potential for quantum information processing \cite{HRB08} and quantum simulation \cite{Schaetz12,BR12}. 
Coulomb crystals with several millions of ions have been observed both in Paul and Penning traps \cite{Walther92a,Walther92b}.
When the inter-ion spacing $a$ is homogeneous, different structural phases can be accessed by tuning the transverse harmonic confinement.
As the trapping frequency $\nu_{t}$ is reduced from high to lower values, the Coulomb crystal undergoes a series of structural phase transitions with phases characterized by linear, zigzag, helicoidal, and more complex structures \cite{Dubin99}. These transitions are generally  of first order,  with the following exception: the transition between the linear ion chain  and the doubly-degenerate zigzag phase, shown in Fig. \ref{fig_ions}(a),  is of second order  \cite{Schiffer93,Piacente04,Fishman08} and occurs at the critical frequency 
\beqa
\nu_{t,c}^2=\frac{7}{2}\zeta(3)\frac{Q^2}{m a^3},
\eeqa
where $\zeta(p)$ is the Riemann-zeta function and $m$ and $Q$ are the mass and charge of the ions, respectively.
A finite-time crossing of this transition is expected to result in the formation of topological defects as described by the KZM \cite{ions1,ions2}, see Fig. \ref{fig_ions}(b).
%
%
%
%

The axial confinement in an ion chain makes the inter-ion spacing spatially dependent, $a=a(x)$, as illustrated in Fig. \ref{fig_ions}(c) and (d). 
Using the local density approximation, away from the chain edges the linear density of ions given by the inverse of the distance between them, $a(x)^{-1}$, is well approximated by the inverted parabola \cite{ions1,ions2}
\begin{equation}
\label{axialdensity}
a(x)^{-1}=\frac{3}{4}\frac{N}{L}\left(1-\frac{x^2}{L^2}\right),
\end{equation}
where $N$ is the number of ions, $L$  is  half the length of the chain, and $x$ the distance from the center.
This leads to a spatial modulation of the critical frequency along the chain, 
\begin{equation}
\nu_{t,c}(x)^2=\frac{7}{2}\zeta(3)\frac{Q^2}{m a(x)^3},
\end{equation}
which ultimately makes the linear to zigzag transition inhomogeneous.
In the thermodynamic limit,  the system obeys an effective time-dependent Ginzburg-Landau equation where the difference
$\nu_t^2-\nu_{t,c}^2$ governs the transition from the high-symmetry phase to the broken symmetry phase \cite{ions1,ions2}.

Consider a  quench of the transverse trap frequency $\nu_t$, 
such that 
\begin{equation}
\nu_t(t)^2=\nu_{t,c}(0)^2-\delta_0^2\frac{t}{\tau_Q}.
\end{equation}
Around the critical point the  transverse frequency can be linearized, $\nu_t\approx\nu_{t,c}(0)-\delta\frac{t}{\tau_Q} $ with $\delta=\delta_0^2/[2\nu_{t,c}(0)]$.
Under such a quench, as a result of the spatial dependence of $\nu_{t,c}(x)$,  the zigzag phase is not formed everywhere at once, and it arises first in the center of the chain.
To account for the formation of kinks it is required to extend the KZM to inhomogeneous scenarios as in section \ref{secIKZM}, see \cite{ions1,ions2} in this context.
The reduced squared-frequency 
\beqa
\varepsilon(x,t)=\frac{\nu_{t,c}(x)^2-\nu_t(t)^2}{\nu_t^{c}(x)^2}
\eeqa
governs the divergence of the correlation length and the relaxation time at the critical point 
\beqa
\xi=\frac{\xi_0}{\sqrt{\varepsilon(x,t)}},\qquad \tau=\frac{\tau_0}{\sqrt{|\varepsilon(x,t)|}}, 
\eeqa
where $\xi_0$ and $\tau_0$ are set by the inter-ion spacing $a(0)=a$ and the inverse of the characteristic Coulomb frequency is given by $\om_0^{-1}=\sqrt{ma^3/Q^2}$.
We have assumed  that the system is underdamped which is the case whenever the dissipation strength $\eta$ induced by laser  cooling satisfies $\eta^3\ll\delta_0^2/\tau_Q$. 
This leads to the critical exponents $\nu=1/2$, and $z=1$.
The front crossing the transition satisfies $\varepsilon(x,t)=0$ and reaches $x$ at time 
\beqa
t_F(x)=\tau_Q\left(1-\frac{\nu_{t,c}(x)^2}{\nu_{t,c}(0)^2}\right).
\eeqa
Relative to this time, it is possible to rewrite the reduced squared-frequency
\beqa
\varepsilon(x,t)=\frac{t-t_F(x)}{\tau_Q(x)},
\eeqa
in terms of the local quench rate 
\beqa
\label{tauqx}
\tau_Q(x)=\tau_Q\frac{\nu_{t,c}(x)^2}{\nu_{t,c}(0)^2}=\tau_Q(1-X^2)^{-3},
\eeqa
where $\nu_{t,c}(x)^2=\nu_{t,c}(0)^2[1-X^2]^3$ and $X=x/L$.
The front velocity reads
\beqa
v_F
\sim\frac{\delta_0^2}{\tau_Q}\left|\frac{d\nu_{t,c}(x)^2}{dx}\right|_{x_F}^{-1}=\frac{L\delta_0^2}{6\nu_{t,c}(0)^2\tau_Q}\frac{1}{|X|}(1-X^2)^{-2}.
\eeqa
The essence of the Inhomogeneous Kibble-Zurek mechanism (IKZM) is the interplay between the  velocity of the front and the sound velocity \cite{KV97,Zurek09}. 
As in  section \ref{secIKZM} (see  Eq. (\ref {hats})), the relevant velocity of perturbations can be estimated to be
\beqa
\hat s =  \frac {\hat \xi} {\hat \tau} = \frac {\xi_0} {\tau_0} \left(\frac {\tau_0} {\tau_Q(x)} \right)^{\frac {\nu(z-1)} {1+\nu z}}= a\om_0.
\eeqa
The last equality holds whenever the dynamic critical exponent $z=1$, as in an underdamped ion chain.

In the IKZM, the condition for kink formation to be possible is given by the inequality 
\beqa
v_{\rm F}(x) >\hat s, 
\eeqa
while the propagation of the pre-selected phase is expected otherwise.
As shown in \cite{DRP11}, it is instructive to study the spatial dependence of the ratio $v_{\rm F}(x)/\hat s$, which as a function of $X=x/L$ turns out to be parametrized by the dimensionless quantity $\mathcal{A}=\frac{L\delta_0^2}{6\nu_{t,c}(0)^2a\om_0\tau_Q}$. 
\begin{figure}
\begin{center}
\psfig{file=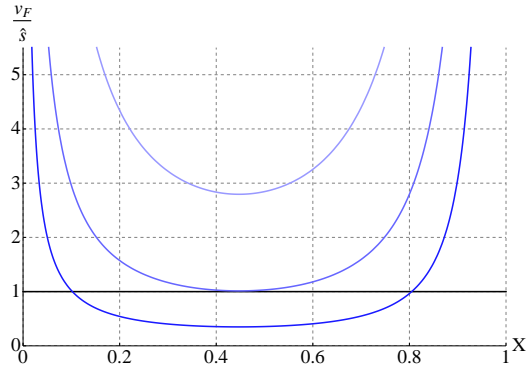,width=0.6\linewidth}
\end{center}
\caption{
Ratio of the front velocity over the second speed of sound as a function of the dimensionless coordinate $X$ measured from the center of the chain ($X=0$) towards an edge ($X=1$, in the Thomas-Fermi approximation). The ratio is symmetric around $X=0$ and only $X>0$ is displayed for clarity. Within the approximation of the IKZM we adopt, formation of defects with KZM densities is only possible there where $v_F/\hat s>1$. The  curves correspond to different values of $\mathcal{A}=\frac{L\delta_0^2}{6\nu_{t,c}(0)^2a\om_0\tau_Q}$ with which color coding increases from light to dark blue taking the values $\mathcal{A}= 0.1,0.289, 0.8$.
Above a critical value $\mathcal{A}_c\approx 0.289$ the homogeneous KZM applies. For $\mathcal{A}<\mathcal{A}_c$ domain formation is expected in two disjoint regions and the inhomogeneous KZM governs the dynamics of defect formation. Disregarding the outer region (where defect losses are dominant), whenever the size of the central region is approximately linear in $\mathcal{A}$, the density of defects scales with a power-law in the quench rate.
}
\label{figvFs}
\end{figure}
Using the Thomas-Fermi approximation for the axial density, Eq. (\ref{axialdensity}),  figure \ref{figvFs} shows that typically $v_{\rm F}(x) >\hat s$ is satisfied in two disconnected regions. 
However, the outer region can be safely disregarded given that kinks possibly formed there are likely to leak out to the linear, outer part of the chain.
A kink experiences a Peierls-Nabarro oscillatory potential whose amplitude diminishes with the transverse amplitude of the zigzag (the order parameter), this is, towards the edges of the chain \cite{ions1,NLkinks2}.
Langevin dynamics simulations show that kinks experiencing a gradient of the zigzag amplitude travel towards the edges of the chain where they disappear.
As a result,  it suffices to consider the central region of the chain  $[-\hat{x},\hat{x}]$ for defect formation. 
Generally $\hat{x}$ has to be found numerically. However, when the defect formation is restricted to a region $\hat{X}\ll1$, then
one can set \cite{ions1}
\beqa
\hat{x}=|\hat{X}|L=\frac{\delta_0^2 L^2}{6\nu_{t,c}(0)^2\tau_Q\hat{v}}+\mathcal{O}(\hat{X}^3).	
\eeqa
Under the assumption $\hat{X}\ll1$, one finds the estimate predicted in \cite{ions1,ions2} for the density of kinks
\beqa
\label{IKZMscaling}
n &\approx & \frac{2\hat{x}}{\hat{\xi}L}=\frac{L}{3\nu_{t,c}(0)^2a^2\om_0^2}\left(\frac{\delta_0^2}{\tau_Q}\right)^{4/3}.
\eeqa
Note that setting $\tau_Q(x)=\tau_Q$ is consistent with $\hat{X}\ll1$.
We note that without restricting to $\hat{X}\ll1$,  there is no reason to expect a power-law scaling, see Ref. \cite{DRP11}.

For sufficiently long quench times, the effective size of a domain set by the KZM length becomes comparable to the (effective) system size, 
$2\hat{x}=2\hat{X}L\sim\hat{\xi}$. In this situation, typically one obtains 0 or 1 defects per realization.
It was pointed out some time ago in the discussion of the winding numbers trapped in loops (see Section \ref{secLoop}), that the scaling with $\tau_Q$ is likely to steepen\cite{KMR00} when the circumference $\cal C$ of the loop becomes less than $\hat \xi$. This prediction has found support in numerical studies of the dispersion of the winding numbers, $\sqrt {\langle {\cal W}^2 \rangle }$\cite{dkzm1,dkzm2,Monaco09}: the doubling of the exponent that  governs the scaling  of the dispersion of $\cal W$ when ${\cal C} \gg \hat \xi$ was seen in the ${\cal C} \ll \hat \xi$ regime, i.e., when ${\cal W}=\pm 1$ is much less likely than the probability of ${\cal W}=0$, as $\sqrt {\langle {\cal W}^2 \rangle }$ is then dominated by the probability of taking a single step in the random walk, while the contribution of ${\cal W} > 1$ is negligible\cite{Zurek13}. 

\begin{figure}
\begin{center}
\psfig{file=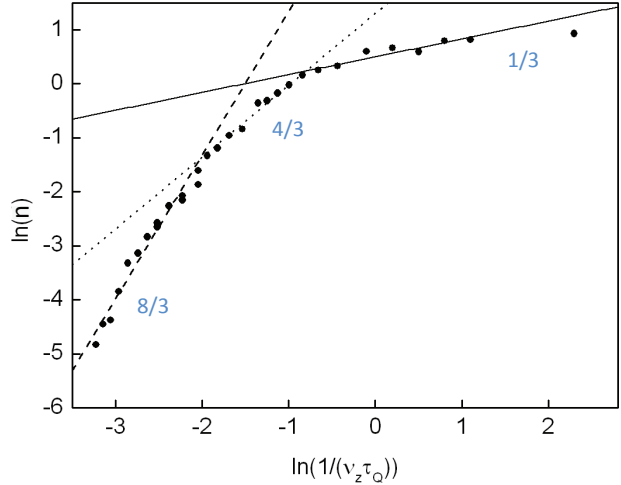,width=0.8\linewidth}
\end{center}
\caption{Scaling of the density of spontaneously formed kinks as a function of the quench rate at which the linear to zigzag structural phase transition is crossed in a trapped Coulomb crystal. 
The simulations are based on  coupled Langevin equations using the experimental parameters in \cite{Tanja13}.  The system is underdamped in the presence of laser cooling.
The plotted lines are guidelines to the eye with slopes predicted by the homogenous KZM ($1/3$), the IKZM ($4/3$), and twice the slope predicted by IKZM.
Adapted from \cite{Tanja13}.
}
\label{fig_multiscaling}
\end{figure}

We have already seen is Section \ref{secLoop} that relating this prediction to experiments requires some care, as doubling of the dispersion of $\cal W$ with $\tau_Q$ in this ${\cal C} \ll \hat \xi$ regime actually implies {\it quadrupling} of the frequency of detections of $|{\cal W}|=1$\cite{Zurek13}, and the frequency is then the obvious observable. In the case of winding number the situation is relatively well understood, at least at the level of simple models. The quadrupling predicted there (and possibly already observed, see Fig. \ref{figFluxon}) is in a sense a product of two doublings. One of them comes about as a consequence of the square root that is related to the random walk of the phase that becomes unnecessary in the case when that random walk has only one step. It is likely relevant only in the case of loops. The second doubling has to do with the change of the character of the question: in small loops the focus is trapping a single defect (and the answer is proportional to the area) while in large loops what matters is the number of pairs intercepted by the  circumference (and the answer is proportional to the circumference, and, hence, to the square root of the area). It is not clear whether at least one doubling survives the transition from loops to open boundary conditions in the case when the size of the system becomes smaller than $\hat \xi$, and excitations becomes rare. Computer experiments with the experimental parameters \cite{Tanja13} are consistent with three regimes: KZM (density scaling with power $\sim \frac 1 3$), followed by IKZM (density scaling steepening to $\sim \frac 4 3$), and, finally -- when the density becomes synonymous with the probability of a single kink -- an even steeper power law that can be interpreted as $\sim \frac 8 3$ of ``doubled'' IKZM, see Fig. \ref{fig_multiscaling}. 
Thus,
assuming a doubling of the IKZM this probability can be estimated to be
\beqa
\label{dikzm}
p_1\sim\left(\frac{2\hat{x}}{\hat{\xi}}\right)^2\sim \frac{L^4}{\nu_{t,c}(0)^4a^4\om_0^4}\left(\frac{\delta_0^2}{\tau_Q}\right)^{8/3}.
\eeqa
Three different experimental groups reported tests of the IKZM in the context of kink formation in trapped ion chains.
Experiments \cite{EH13,Tanja13} followed closely the proposal in \cite{ions1,ions2}, where critical dynamics was driven by a finite-rate decompression of the transverse confinement.
The experiment at Mainz \cite{Ulm13} used instead a compression along the axial direction.
The system sizes  and the accessible quench rates in these experiments correspond precisely to the onset of adiabatic dynamics, 
where $\{0,1\}$ defects are observed per realization \cite{Ulm13,Tanja13}.
 The experiment at Simon Fraser University \cite{EH13}  reported  broader kink number  distributions but the presence of  substantial kink losses prevented  testing any signature of universality in the dynamics of kink formation. See Table \ref{ion_exps_table} for a summary of  these experiments. The results of \cite{Ulm13,Tanja13} suggest an agreement with (\ref{dikzm}). 
%
\begin{table}[h]
\tbl{Experimental results on the topological defect formation in ion Coulomb crystals \cite{Ulm13,Tanja13,EH13}. 
Data was fitted to a power-law in the quench rate $\tau_Q$ of the form $n\propto \tau_Q^{-\alpha }$.}
{\begin{tabular}{@{}cccc@{}}\toprule
Group                                 &    Number of ions      &  Kink number   &    Fitted exponent  $\alpha$             \\
\colrule
                                          &                           &                                     &                                                 \\
Mainz University   \cite{Ulm13}             &     $16$     &        $\{0,1\} $              &           $2.68\pm 0.06$           \\
                                           &                          &                                     &                                              \\
PTB  \cite{Tanja13}                                   &    $29\pm2$                       &        $\{0,1\} $             &           $2.7\pm 0.3$               \\

                                          &                           &                                     &                                                 \\
Simon Fraser University  \cite{EH13}   &     $42\pm1$     &        $ \{0,2\} $              &            $  3.3\pm0.2$                      \\[1ex]
\botrule
\end{tabular}
}
\label{ion_exps_table}
\end{table}
%
%

There are obvious concerns about the extent to which the limited data behind Table 
\ref{ion_exps_table}  can be regarded as a verification of the KZM (e.g., the restricted range of quench rates in each regime, the losses of defects, etc.). However, over and above such experimental issues there are two concerns related to the applicability of IKZM theory to the ion chains of the size $\hat X \ll \hat \xi$ accessible so far in the experiments\cite{Ulm13,Tanja13,EH13}. 

Our discussion in Section \ref{secIKZM} was based on the idea (put forward in the analysis of soliton formation\cite{Zurek09}) that a system 
in an anisotropic harmonic trap can be cleanly divided into regions where the phase transition front velocity is faster (or slower) than the relevant speed of sound. As a consequence, one can distinguish regions where the homogeneous KZM holds (or does not) and defects are created with the local density set by $\hat \xi$ (or not at all).

This sharp division is the key assumption underlying the IKZM. However, in computer simulations and analytic studies the transition between the ``homogeneous KZM'' and ``no defects at all'' is not completely sharp, and it seems unlikely (e.g., in view of the behavior of the order parameter in the presence of the gradients\cite{ZD08}) that it could be less than $\hat \xi$. Thus, the applicability of the IKZM scalings to the ion chains where $\hat X \ll \hat \xi$ can be questioned at least in the experiments with rather small systems\cite{Ulm13,Tanja13,EH13}, as the limits on the integral in Eq. (\ref{dikzm}) are not well defined. 

The above concern may apply to the IKZM in all small systems. It appears in addition to the difficulties involved in testing the KZM in many-body systems of modest size, where the scaling in the near-critical regime may not have converged to the values of critical exponents that determine the universality class. This concern can be of course addressed by carrying out suitable equilibrium measurements to verify $\nu$ and $z$, and determining that they extend over the range relevant for the KZM. 

The KZM is a way to employ equilibrium scalings in predictions of the consequences of non-equilibrium quenches. Checking if the system in question follows the behavior predicted for its equilibrium universality class seems like a prudent first step when the systems is of modest size and especially if it is in a trap or any other setting that can potentially complicate its behavior.

The other way to address such concerns is to work with large homogeneous systems. In case of ion traps this is not out of the question: large ``racetrack'' traps are in principle possible \cite{Walther92a} and could be used to study phase transitions and symmetry breaking in ion Coulomb crystals in a setting where the homogeneous KZM could be tested\cite{Nigmatullin11}.



\subsection{Prospects of ground-state cooling of ion chains}

In the theory and experiments just discussed, the ion chain is hot enough so that thermal fluctuations dominate over quantum fluctuations, and a classical description applies.

The prospects of achieving ground-state cooling, while experimentally challenging, might pave the way to study the dynamics of  quantum phase transitions in ion traps and similar settings.
The equilibrium and dynamic properties of the quantum linear to zigzag structural transition have been investigated \cite{QPTions0,QPTions1,QPTions2,QPTions3,QPTions4}.

Accessing the quantum regime would also pave the way to the experimental realization of topological Schr\"odinger cats, nonlocal quantum superpositions of 
conflicting choices of the broken symmetry or quantum phases of matter\cite{Qkinks1}. Superposition of macroscopic states have been also explored in the context of ion Coulomb crystals \cite{Qkinks2} and magnetic fields coupled to quantum many-body systems\cite{RZD12}. 
Quantum solitons are expected to exhibit long coherence times in the presence of cooling in the Doppler limit, and can be manipulated thanks to the 
spectral properties of the internal modes, which have  been proposed as carriers of  quantum information \cite{Qkinks3}.
As a test-bed for entanglement generation\cite{Cincio} and the subsequent decoherence\cite{Qkinks4}, the creation of quantum structural defects might shed new light 
on fundamental issues concerning the relation between decoherence and critical phenomena \cite{DQZ11}.


\section{Soliton formation in  Bose-Einstein condensation}
\label{secSoliton}

One of us has suggested the finite-rate Bose-Einstein condensation of a thermal cloud in an elongated trap as an inhomogeneous test-bed for the KZM \cite{Zurek09}.
The inhomogeneity of the trap plays the key role in re-setting the dependence between the quench rate and the number of defects---solitons in a BEC ``cigar'' (see Fig \ref{Cigar}). The study of Bose-Einstein condensation in a harmonic trap motivated the development of the IKZM theory we have presented in Section \ref{secIKZM}.

\begin{figure}
\begin{center}
\psfig{file=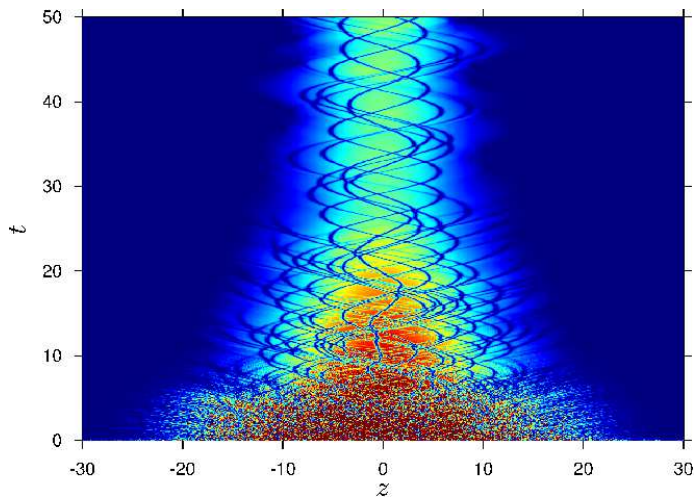,width=0.8\linewidth}
\end{center}
\caption{Time evolution of the density profile in a single realization of the cooling ramp simulated with the classical field method \cite{WDGR11}. The initial state is chosen from a canonical ensemble above the critical temperature for Bose-Einstein condensation. Evaporative cooling is simulated by a linear ramp  of the axial trapping potential using a one-dimensional generalized Gross-Pitaeveskii equation.
Courtesy of E. Witkowska and P. Deuar. 
}
\label{fig_WD}
\end{figure}

This proposal has recently been realized in the laboratory at the BEC center in Trento \cite{Lamporesi13}. 
The basic idea is that as a thermal cloud of atomic vapor undergoes  evaporative cooling through the critical temperature for Bose-Einstein condensation, 
different regions of the newborn condensate pick up a different condensate phase. 
When two neighboring regions merge, the mismatch in the phase of the condensate wavefunction 
acts as a seed for the formation of a phase jump and the corresponding density dip:  a gray soliton is spontaneously formed.
Numerical simulations based on the stochastic Gross-Pitaevskii equation indicate that this scenario in a homogeneous cloud 
is well-described by the Kibble-Zurek mechanism \cite{DZ10,WDGR11}.
As an instance of a single realization, figure \ref{fig_WD} shows the time evolution of the density profile of a newborn condensate following an evaporative cooling ramp.
From the trajectory of the density dips, it is apparent that these excitations constitute spontaneously formed solitons. Note however that in thermal equilibrium 
solitons are as well expected to be formed \cite{Karpiuk12}.

In harmonic traps, the dynamics of Bose-Einstein condensation is more complex due to the inhomogeneous nature of the system. 
The atomic cloud is trapped in an anisotropic three dimensional harmonic confinement $U(r,z)$ with a cigar-shape, characterized by an axial frequency $\om$ and a transverse 
one $\om_{\perp}$ ($\om_{\perp}>\om$).
 The density of the cloud is highest at the center of the trap. Disregarding the transverse degrees of freedom, one 
can use the local density approximation to estimate the value of the critical temperature based on the Einstein equation with an axial dependence,
\beqa
T_c(z)=\frac{2\pi\hbar^2}{m k_B}\bigg[\frac{\rho(r=0,z)}{\zeta(3/2)}\bigg]^{\frac{2}{3}},
\eeqa
which is obtained by replacing the constant density $\rho$ by $\rho(r=0,z)$ in the expression for the critical temperature in a homogeneous system.

 In the experiment\cite{Lamporesi13}, a radio-frequency knife is used to force the evaporation of the cloud, by flipping the atomic spin from a trapped state to an untrapped state.
Atoms with a potential energy $U_{RF}=h\nu_{RF}$ measured with respect to the bottom of the trap are forced to evaporate. 
The resulting axial temperature profile is given by 
\beqa
T(z)=\frac{U_{RF}-U(r,0,z)}{\eta k_B}.
\eeqa
\begin{figure}
\begin{center}
\psfig{file=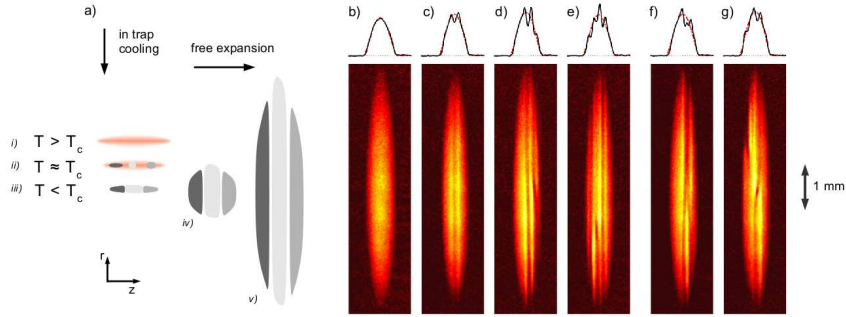,width=\linewidth}
\end{center}
\caption{Spontaneous soliton formation under forced evaporative cooling of a cigar-shaped atomic cloud \cite{Lamporesi13}.
a) As the temperature (i-iii) is decreased below its critical value for Bose-Einstein condensation, causally disconnected regions of the newborn condensate 
cloud pick up different phases of the condensate wavefunction, and the subsequent dynamics leads to the formation of solitons. Under time of flight (iv-v), an initial cigar-shaped cloud expands mainly along the transverse direction.   b)-g) Snapshots of the density profile after time of flight of clouds containing $0,1,2,3,$ solitons and  two instances exhibiting bending of the soliton in the transverse direction. Fits to the self-similar expansion of the Thomas-Fermi density profile (red line) are compared with integrated density profiles of the central region of the cloud (black line).
Reprinted by permission from Macmillan Publishers Ltd: Nature Physics Ref. \cite{Lamporesi13}, copyright (2013).
}
\label{fig_solitons}
\end{figure}

In-situ optical imaging of solitons is challenging due to the smallness of the typical values of the healing length, which sets the width of the soliton. As a result, experimental measurements often resort to time-of-flight (TOF) imaging which magnifies the  size of the cloud, see Fig. \ref{fig_solitons}.  When interactions can be disregarded during TOF, the dynamics is essentially ballistic, and the evolution of local correlations such as the density profile is rather trivial.
This is the case  when the anisotropy of the trap is not too large. 
When the confined cloud acquires an effectively one-dimensional character, no true BEC is possible \cite{DSW00}, and the presence of phase fluctuations in the trapped superfluid severely distorts the TOF dynamics.
Counting of solitons by imaging the density profile of the cloud after time of flight is then hindered by the fact that phase fluctuations map to density fluctuations as suggested in \cite{ripples}, 
and experimentally demonstrated in \cite{g2exp}. This mapping precludes establishing a correspondence between the density of fringes in the expanded cloud 
and the initial number of solitons, formed spontaneously as a result of evaporative cooling, presumably described by the IKZM.
 As a consequence, the anisotropy should be low enough to minimize phase fluctuations associated with the low-dimensional character of the cloud. At the same time, it should be high enough to reduce the role of the instability and decay mechanisms of solitons in elongated three-dimensional clouds, such as the snake instability.
This is the regime of relevance to the experiment \cite{Lamporesi13}. Indeed, TOF snapshots  f) and g) in figure \ref{fig_solitons} suggest bending of the soliton along the transverse degree of freedom in the trapped cloud. 
The time of flight achieved to image the system were particularly large since TOF was assisted by levitation.
 We point out that as an alternative, shortcuts to adiabatic expansions \cite{delcampo13,STA1,STA2}
can be used with similar outcomes \cite{Schaff1,Schaff2}.  

%
\begin{table}[h]
\tbl{Power laws predicted by the Kibble-Zurek mechanism for the number of solitons $N_s$ as a function of the cooling rate $1/\tau_Q$  induced by forced evaporation through the critical temperature for Bose-Einstein condensation of a cigar-shaped atomic cloud. The exponent $\alpha$ of the power-law $N_s\sim\tau_Q^{-\alpha}$ is shown
for different critical exponents ($\nu$, $z$) and trapping potentials.}
{\begin{tabular}{@{}cccc@{}}\toprule
Critical exponents         &    Homogeneous    system       &    Harmonic    trap                \\
\colrule
             &                          &                                          \\
Arbitrary    ($\nu$, $z$)
             &  $\frac{\nu}{1+\nu z}$  &  $\frac{1+2\nu}{1+\nu z} $    \\
             &                          &                                          \\
Mean-field theory ($\nu=\frac{1}{2}$, $z=2$)
             &    $\frac{1}{4}$         &      $1$                         \\
             &                          &                                                  \\
Experiments/F model\cite{HH77}   ($\nu=\frac{2}{3}$, $z=\frac{3}{2}$)
             &     $\frac{1}{3}$        &      $\frac{7}{6}$                \\[1ex]
\botrule
\end{tabular}
}
\label{powerlawstable}
\end{table}
%

The upshot of the counting statistics was a power-law dependence of the mean number of solitons $N_s$ in the cooling rate $1/\tau_Q$ induced by forced evaporation, i.e., $N_s\sim\tau_Q^{-\alpha}$. The power-law exponent resulting from a fit to the experimental data was found to be $\alpha=1.38\pm0.06$, which clearly deviates from the homogeneous KZM exponent  and suggests a possible  agreement  with the IKZM, as illustrated in Table \ref{powerlawstable}. This exponent was shown to be robust against variations of the 
mean atom number of the newborn condensate (which was varied from 4 to 25 million atoms), an indication of universal behavior.  

 The KZM as well as the IKZM can be used as a tool to determine critical exponents \cite{DZ10}, an application of interest in Bose-Einstein condensation.
The experiment \cite{Donner07} reported deviations from mean-field behavior, $\nu=1/2$. The BEC transition is believed to belong to the static 3D XY universality class, for which $\nu = 0.6717(1)$ according to the theoretical estimate in \cite{3DXY}. Let us ignore for the moment possible systematic experimental errors, and assume that the power-law exponent $\alpha$ measured in the Trento experiment is actually given by the IKZM, $\alpha=\frac{1+2\nu}{1+\nu z} $.
In principle, one can extract the value of the dynamical critical exponent $z \simeq 1.04\pm0.11$, which has not been directly measured so far in experiments.
This would rule out both the mean-field value  $z=2$ (by  8-$\sigma$) as well as the F model $z=\frac 3 2$ value\cite{HH77} (by  4-$\sigma$),
when the experimental data is taken ``at the face value'', 
i.e., without accounting for the possible systematic effects, such as decay of solitons (that may steepen the dependence),  as well as the fact that the number of solitons created in the trap is of order 1 (which may result in steepening of the dependence of their number on the quench rate we have discussed in the previous section). There is also a concern signalled by the authors of the experiment that the axial temperature in the BEC cloud is not uniform, which may further modify the scaling. Experiments in larger traps that lead to more solitons would be helpful in addressing this concern. 


Numerical simulations using the classical field method \cite{WDGR11} lead to similar power-law exponents which would agree with the IKZM, 
but are based on a model of evaporation along the axial direction which is not applicable to the experiments.


\section{Vortex formation in a newborn Bose-Einstein condensate}
\label{secVortex}

The observation of spontaneous soliton formation during Bose-Einstein condensation was actually preceded by analogous experiments in pancake-shaped  atomic clouds \cite{Anderson08}.
The process is fairly similar. Consider a thermal quench between an initial temperature $T_i$
and a final value  $T_f$
which is linear in time, i.e.,  $T(t) = T_i-t\frac{T_i-T_f}{\tau}$.
During the cooling of the atomic cloud below the critical temperature for Bose-Einstein condensation, coherent patches are created where 
the phase of the condensate wavefunction is chosen independently and is approximately constant. When these different regions merge, there is a chance for the phase to accumulate along a close loop  in (physical) space surrounding  a given point. 
Indeed, it was shown experimentally that the explicit merging of three independent BEC clouds results in the formation of a vortex with a  certain probability given by the geodesic rule \cite{Anderson07}.
Consider the expectation value of the order parameter $\la \hat \psi\ra_0=|\psi|e^{i\theta(x)}$.
The phase accumulated around a loop should be an integer modulo $2\pi$ and can be characterized by the winding number
$\mathcal{W}=\frac{1}{2\pi}\oint\partial \theta$.
The fundamental (or first) homotopy group is indeed given by the ring of integers, $\pi_1(U(1)/\{1\})=\pi_1(S^1)=\mathbb{Z}$ (see Appendix \ref{secSSB}).
Whenever $|\mathcal{W}|\geq1$ a line defect, string or vortex is formed.
In a homogeneous system, or for fast enough quenches in a trapped cloud, 
the density of vortices is expected to be given by Eq. (\ref{dhkzm}), i.e., the homogenous KZM scaling,
\beqa
n=\frac{1}{f^2\hat{\xi}^2}=\frac{1}{f^2\xi_0^2}\left(\frac{\tau_02\delta}{\tau T_c(0)}\right)^{\frac{2\nu}{1+\nu z}},
\eeqa
while when the influence of the harmonic confinement is taken into account\cite{Zurek09,DRP11} (now in an approximately 2D BEC ``pancake'',  rather than a``cigar''), 
\beqa
n\propto
\left(\frac{\tau_02\delta}{\tau T_c(0)}\right)^{\frac{2(1+2\nu)}{1+\nu z}},
\eeqa
as discussed in Section \ref{secIKZM}.
The experiment \cite{Anderson08}  reported as well the spontaneous vortex formation in a toroidal trap. 
This geometry offers the opportunity to explore a scenario where the condensation is inhomogeneous in the transverse degree of freedom and homogeneous in the toroidal direction.
The density of vortices is predicted to scale then as \cite{DRP11}
\beqa
n\propto
\left(\frac{\tau_02\delta}{\tau T_c(0)}\right)^{\frac{1+3\nu}{1+\nu z}}.
\eeqa
The accurate experimental determination of the power-law exponents in any pair of these three scenarios, which remain untested to-date, 
would allow the independent determination of the critical exponents $\nu$ and $z$.


\section{Mott Insulator to superfluid transition}


A natural testing ground for the KZM in quantum phase transitions is the transition between a Mott insulator (MI) and a superfluid (SF) phase,  
exhibited by the Bose-Hubbard model \cite{Fisher89,Jaksch98} and accessible to quantum simulation based 
on ultracold gases in optical lattices \cite{Greiner02,Bakr10}.
The Hamiltonian of the system is
\beqa
\hat{\mathcal{H}}_{BH}=-\sum_{\la i,j\ra}J_{ij}(\hat{b}_i^{\dag}\hat{b}_j+h.c.)+\frac{U}{2}\sum_i\hat{n}_i(\hat{n}_i-1)-\mu\sum_i\hat{n}_i
\eeqa
where $\mu$ is the chemical potential, the tunneling matrix element for an atom to hop from site $i$ to $j$ is given by $J_{ij}$, $\hat{b}_i$ and $\hat{b}_i^{\dag}$ are respectively the annihilation and creation operators, and $U$ is the on-site interaction.
We shall simplify the discussion by setting $J_{ij}=J$. In actual experiments the optical lattice implementing $\hat{\mathcal{H}}_{BH}$ is usually contained in a harmonic trap that will result in, e.g., the couplings $J_{ij}$ that are spatially dependent. 

In the limit $J/U\ll1$ the homogeneous system is in the Mott insulator phase whenever  the filling is commensurate with $n$ atoms per site.
The many-body state takes the form $|MI\ra\propto\prod_{i=1}^M(\hat{b}_i^{\dag})^n|0\ra$. 
The opposite limit $J/U\gg1$ corresponds to the superfluid phase with a many-body state of the form $|SF\ra\propto\left(\sum_{i=1}^M\hat{b}_i^{\dag}\right)^N|0\ra$, characterized by phase coherence and large fluctuations in the number of particles per site.
\begin{figure}
\begin{center}
\psfig{file=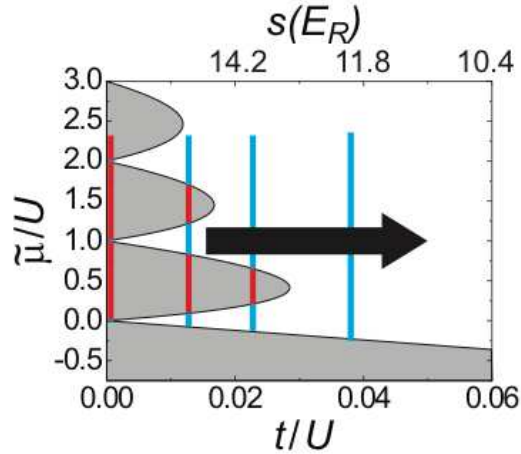,width=0.7\linewidth}
\end{center}
\caption{Phase diagram of the Bose-Hubbard model illustrate the boundary between  the Mott insulator  and superfluid phases. The vertical lines range over the densities and effective chemical potential $\tilde{\mu}$ sampled in a single experimental realization. The quench is driven by a fast modulation of $J=\mathlarger{\mathlarger{\mathlarger{t}}}$ in the direction of the black arrow.
From {\it Chen et al.} \cite{DeMarco11}. Copyright 2011 American Physical Society.
}
\label{fig_MISF}
\end{figure}
The transition from the MI to the SF can be driven by increasing the relative coupling $(J/U)(t)$ in a finite time $\tau_Q$. 

The buildup of correlations induced by such a quench has been analyzed in a series of works \cite{RUXF06,Cucchietti07,DMZ08,Dziarmaga10,DTZ12,TDZ13}. In a 3D experimental realization of the Bose-Hubbard model, an interesting systematic study of the amount of excitations and energy produced during a quench as a function of the quench rate was undertaken, and found a power-law dependence suggesting a KZM-like behavior \cite{DeMarco11}. 
The amount of excitations was estimated by comparing the density profile after time of flight (TOF) for  an initial cloud in the ground state ($n_0(x,y;t)$) and an out-of-equilibrium cloud prepared by a quench ($n(x,y;t)$). The following quantity was used
\beqa
\tilde{\chi}^2=c\; \mathcal{N}\int dx dy\frac{[n(x,y;t)-n_0(x,y;t)]^2}{n_0(x,y;t)}
\eeqa
where $\mathcal{N}=\int dx dy n(x,y;t)$ and $c$ is a constant chosen in agreement with numerical simulations.
 The dependence of $\tilde{\chi}^2$ was fitted to a power-law $\tilde{\chi}^2\propto\tau_Q^{-\alpha}$ with $\alpha=0.31\pm 0.03$.
The kinetic energy $K=m\la r^2\ra/(2t_{TOF})$, where  $t_{TOF}$ is the time of flight and $\la r^2\ra=N\int dx dy (x^2+y^2)n(x,y;t)$,
was as well analyzed and fitted to a power law with a comparable exponent  $\alpha=0.32\pm 0.02.$


According to Chen et al., these exponents deviate from theoretical predictions for a 3D homogeneous transition.
   They studied the dependence on the quench rate of the density fluctuations characterized by $\tilde{\chi}^2$ and the kinetic energy, and compared it with a power-law with exponent $\alpha=\frac{3\nu}{\nu z+1}$, i.e., which would result when the fluctuations and excess of kinetic energy are the same in each domain of size $\hat \xi$.
For $\nu=1/2$ and $z=2$ this would lead to $\alpha=3/4$\cite{DeMarco11}.

\begin{figure}
\begin{center}
\psfig{file=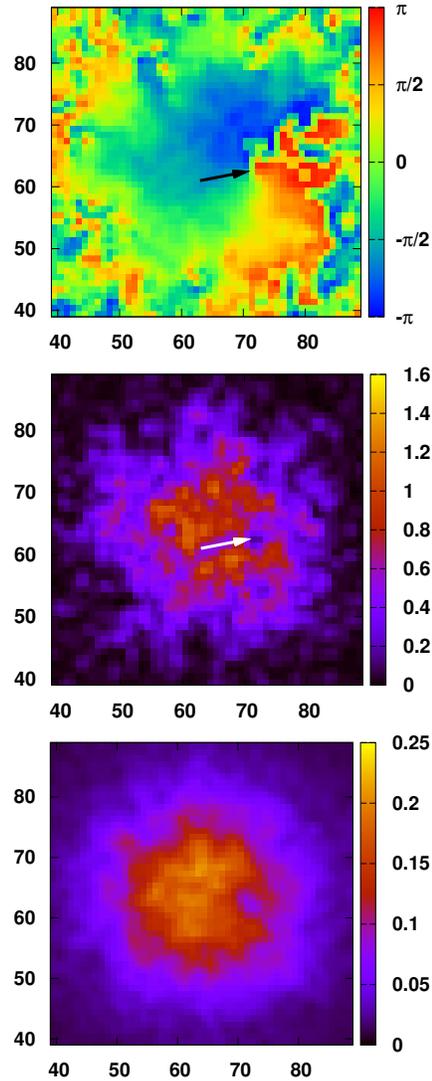,width=0.5\linewidth}
\end{center}
\caption{Simulation of a linear quench from  the Mott insulator to superfluid phase in a three dimensional trapped atomic cloud using a discrete Gross-
Pitaevskii equation. The top and medium panel correspond to phase and density in a $x-y$ cross section across the center of the trap. The integrated density along the perpendicular direction is shown in the lower panel. The position of a vortex is indicated by an arrow. As noted in the discussion, the scaling of, e.g., the kinetic energy with the quench rate is similar to what was observed on the experiment of Chen et al. This coincidence  between the effectively classical simulation (attained in the large occupancy per site limit) and the experiment (that had about 3 atoms per site)  may be a accidental, or may be due to the fact that BEC in the experiment was decohered\cite{DDZ} by, e.g., the finite temperature effects and the significant normal fraction. From {\it Dziarmaga et al.} \cite{DTZ12}.  Copyright 2012 American Physical Society.
}
\label{fig_MISF2}
\end{figure}
We shall not attempt to justify these assertions here, nor shall  dispute them. Clearly,  if one were to accept validity of the homogeneous KZM for this relatively modestly sized inhomogeneous system of about $1.6\times 10^5$ atoms (a BEC-filled optical lattice confined to an inhomogeneous sphere of radius of only $\sim$10 sites, each occupied by $\sim 3$ atoms), more detailed analysis would be useful to explore the conjecture that the applied measures of the degree of excitation imparted by the quench result in the behavior described simply by the same kinetic energy and same departure from homogeneity in each $\sim \hat \xi^{-3}$ volume, which is what the ansatz\cite{DeMarco11} described above suggests. Moreover, 
the experimental conditions hindered a direct connection with this KZM estimate. 
In particular, the inhomogeneous character of the system induced by the presence of an external harmonic trap results in an initial state with  Mott insulating layers of different filling factors being separated by (presumably phase coherent) superfluid layers, see Fig. \ref{fig_MISF}. In addition, the phase boundary is crossed at a range of densities, with different MI layers (corresponding to different lobes in the phase diagram)  crossing the phase transition at different times. The analysis of this scenario is substantially more complicated than the exposition of the IKZM discussed in Section \ref{secIKZM}, where power-law behaviors are expected for a single critical front. The finite-temperature of the initial cloud in \cite{DeMarco11} has also not been taken into account in KZM studies applied to this transition.

We note that the numerical simulations\cite{DTZ12} in the limit of large occupancy per site yield power laws, e.g., for the kinetic energy, that are similar to those observed in the experiment. Moreover, topological defects appear in the superfluid left after the Mott insulator -- superfluid phase transition, see figure \ref{fig_MISF2}. It is again far from clear whether this coincidence of scalings is significant. In the large occupancy regime the system is effectively classical (which is what makes the computer simulation possible in the first place). On the other hand, in the actual experiment there was a substantial ($\sim 10\%$) normal fraction and non-negligible temperature. That combination may cause decoherence\cite{DDZ} and, hence, force a quantum many-body system to behave in an effectively classical manner.


 \section{Summary and Outlook}

The Kibble-Zurek mechanism reviewed here is based on the combination of two key ideas. The seminal observation of Tom Kibble\cite{Kibble76,Kibble80} made it clear that, at least in the cosmological context, phase transitions expected to occur as the Universe cools soon  after the Big Bang will result in a mosaic of domains of the size close to the Hubble radius at the time of the transition. This is simply a consequence of relativistic causality---domains are forced to break symmetry independently, and, hence, at random. Moreover, when the resulting homotopy group is nontrivial, phase ordering cannot completely smooth out the post-transition configurations of the order parameter, as the random choices of broken symmetry lead to irreconcilable differences that  crystalize into topological defects.

In the second order phase transitions encountered in the laboratory relativistic causality does not yield useful limits, but the speed of light can be effectively replaced by the relevant speed of sound\cite{Zurek85,Zurek93,Zurek96} leading to an estimate of the size $\hat \xi$ of the domains that can consult on how to break the symmetry, and, hence,  that can choose to break symmetry  more or less in unison. The resulting density of topological defects and other excitations, left behind by phase transitions induced at a finite speed, depends on the interplay of the quench rate (the rate at which the critical point is traversed) and critical slowing down (the rate with which systems can adjust), and as a result, on the universality class of the transition. The scaling of $\hat \xi$ with the quench rate (reflected in the density of defects left behind by the transition) can be investigated in the laboratory. 

Experiments testing KZM scaling were the focus of our review. The Kibble-Zurek scaling was also tested in classical and quantum phase transitions in a variety of computer experiments\cite{
LagunaZ1,LagunaZ2,YatesZ,Hindmarsh,Stephens,KibbleRajantie,DLZ99,Caneva07,Pellegrini08,Monaco09,Bermudez09,DZ10,dkzm1,dkzm2,WDGR11,Deng11,SZD1,SZD2,Cugliandolo11,Vacanti12,Brand13,SWDM13,SKU13,Miranda13b} and analytical works \cite{Damski05,Dziarmaga05,Polkovnikov05,Lamacraft07,USF07,chandran1,chandran2},  and found to hold essentially whenever it was expected to apply. Laboratory experiments are, of course, more difficult. Above all, it is hard to vary the quench rate over several orders of magnitude (needed  to detect the fractional power laws predicted for $\hat \xi$ as a function of $\tau_Q$) while avoiding effects that can either suppress generation of topological defects (e.g., inhomogeneities) or result in formation of defects in processes (e.g., convection in superfluids) independent of KZM  that could obscure KZM-predicted scaling. Moreover, defects formed in the course of the transition can annihilate during the phase ordering that follows the transition.                                                                                                                                                                                                                                                                                

A brief summary of the present day ``experimental KZM landscape'' is that there are now several experiments that have found, in various systems, results consistent with KZM scalings. However, all of them require at present caveats and additional assumptions for interpretation.

Switching between non-equilibrium steady states provided early evidence for KZM scaling\cite{Ducci99}. Nonetheless, subsequent experiments as well as numerics indicated that in such situations where the renormalization theory cannot be invoked KZM scalings may be only an approximation or not apply\cite{Ashcroft13}.  Still, such efforts have led to the earliest experimental indications of the KZM scaling, and may offer intriguing opportunities for extension of KZM to transitions that are not described by renormalization or even by partial differential equations.

Trapping of flux quanta in tunnel Josephson junctions yielded scaling that appears to be reliable, but the detected exponent of $\sim 0.5$ was twice what was initially expected. That expectation was based on the prediction of the doubling of the power law for large winding numbers\cite{KMR00}. Recent analysis\cite{Zurek13} of the winding numbers in the case of small loops (${\cal C} \ll \hat \xi$) indicates that, while one would indeed expect the exponent that governs the {\it dispersion} of $\cal W$ to double in the regime where $|{\cal W}|>1$ is vanishingly unlikely, the {\it frequency} of trapping of $|{\cal W}|=1$ scales with four times the power predicted for large $|\cal W|$. This suggests that the KZM accounts for the experimental results.
This quadrupling may be also relevant for small superconducting loops, where the observed frequency of trapping a flux quantum scales with the exponent $\simeq 0.62 \pm 0.15$\cite{Monaco09}, consistent with 0.5 seen in tunnel Josephson junctions\cite{Monaco08} (although possibly suggestive of $\frac 2 3$, which is---one might be even tempted to speculate---four times $\frac 1 6$, the exponent expected for the scaling of typical winding numbers trapped by large loops for $\nu= \frac 1 2$ and $z=1$ or a superfluid with $\nu= \frac 2 3$ and $z= \frac 3 2$, where $\hat \xi \sim \tau_Q^{\frac 1 3}$). 

We note that all these discussions of doubling and quadrupling ignore the role of the magnetic field, which, as was pointed out in the case of loops\cite{Zurek96} and demonstrated much more clearly in 2D systems with the help of numerical simulations\cite{Hindmarsh,Stephens,KibbleRajantie} may play a significant role in flux trapping and defect formation in systems with local gauge invariance.

Defect formation in multiferroics is a new frontier. Experiment in ErMnO$_3$ yields a compelling power law\cite{Chae12}, but its interpretation in terms of KZM depends on the nature of the critical region of the transition  that is inaccessible to, e.g., susceptibility measurements, as a result of the high critical temperature. Still, theoretical analysis\cite{Griffin12} based on the 3D XY model yields an impressive agreement of KZM with the experiments. Nevertheless, a more precise determination of the exponent that governs the power law scaling that would clearly establish the connection with the $\nu$ and $z$ predicted for the 3D XY universality class would be welcome: it would amount to the first experimental confirmation of KZM scaling in a setting  that is not mean field. 


Generalization of the KZM to inhomogeneous systems is usually needed to interpret experiments in harmonically trapped ions and BEC's. Formation of kinks in ion Coulomb crystals\cite{Tanja13,Ulm13,EH13} and solitons in Bose-Einstein condensation\cite{Lamporesi13} has been recently reported. It has been argued in both ion crystals and BECs that the data are consistent with the KZM when one recognizes both the consequences of inhomogeneity and (in the case of the ion chains) small size of the system.   Dependence of the conclusions about scaling on these additional assumptions complicates the interpretation especially in the case of kinks in ion chains, but the results are consistent with the suitably modified versions of the KZM. 

In the case of BEC solitons the measured power laws are close to the analytic predictions\cite{Zurek09}, and the remaining discrepancy may be due to the difference between the simpliefied effect of the harmonic trap analyzed theoretically \cite{Zurek09} and the experimental reality\cite{Lamporesi13}. Indeed, corrections to  a power-law scaling can be expected in inhomogeneous systems \cite{DRP11,DKZ13}. Additional experimental results and theoretical as well as numerical efforts would certainly be useful. 

The presence of vortices in a newborn BEC has also been detected\cite{Anderson07,Anderson08}. They have appeared presumably as the result of the KZM. However, obtaining reliable power laws in this case is even harder. The experiments are carried out, e. g., in approximately 2D BEC pancakes, so inhomogeneities and small systems sizes will play a role and complicate the analysis.

An experiment, that has not been completed as yet but is under way in the group of Markus Oberthaler\cite{talkUCSB}, probes a quantum miscibility-immiscibility phase transition in a Bose-Einstein condensate\cite{Timmermans98}. Numerical simulations in an effectively 1D toroidal trap yield scalings of the size of domains of the two hyperfine BEC states that are in good agreement with the KZM prediction\cite{SZD1,SZD2}. In the harmonic trap (where the actual experiments will likely take place) inhomogeneity modifies domain sizes in a way that influences the observed power law, again complicating direct comparisons with the KZM predictions, although numerical   simulations may help.

Experimental investigations of the quantum KZM (exemplified by the miscibility-immiscibility transition) are only beginning. Experiments to date (e.g., related to the Bose-Hubbard model\cite{DeMarco11}) suffer from complications caused by the inhomogeneities and small system sizes. In view of the rather complicated phase diagram of the Bose-Hubbard model, inhomogeneities make critical exponents relevant for the KZM scaling difficult to infer. Moreover, computer simulation of the quantum Bose-Hubbard model are difficult, as systems of sizes large enough to hope for a suitably well defined quantum phase transition are also large enough to be essentially out of reach of present day computers. One can study larger systems only in the limit where they become effectively classical---when the number $N$ of atoms per site is large\cite{DTZ12,TDZ13}. When one compares results of such simulations with the data obtained in experiments, there are no obvious discrepancies that cannot be blamed on inhomogeneity or finite size, but this rather tentative conclusion (based on the comparison of a classical simulation to a quantum many-body system) is unsatisfying and it certainly leaves plenty of room for improvement.
															
This cautious assessment of the present status of the experiments on the dynamics of quantum phase transitions is likely to be revised in the near future. 
Moreover, quantum phase transitions in the Ising model (which is much better understood theoretically) may be eventually implemented (e.g., by emulating its dynamics\cite{Monroe11,SPS12}) in suitably large systems. This would be interesting not just because of the implications for the KZM, but because one could then study non-local superpositions of topological defects---``topological Schr\"odinger cats'' or ``Schr\"odinger kinks''\cite{Qkinks1}---as well as probe the possibility of a quantum superposition of distinct phases of matter\cite{RZD12}. Last, but not least, there are examples (e.g., quantum Ising model) where the KZM predicts the range of entanglement in the post-transition state\cite{Cincio} and even the effectiveness of the system undergoing second order phase transition as an environment responsible for decoherence\cite{DQZ11}.

Kibble-Zurek mechanism employs {\it equilibrium} behavior of the system to predict  non-equilibrium consequences of the  dynamics of symmetry breaking. When we compare the first experimental tests of the KZM with some of the recent experiments, one important difference stands out: in the pioneering tests of the KZM the equilibrium behavior of the systems in the vicinity of the critical point was generally very well known as a result of earlier measurements, so the scaling predicted by the universality class was beyond doubt. This is often not the case in the recent KZM-inspired experiments of, for example, quantum phase transitions in optical lattices. 
It would seem prudent to test equilibrium of the actual system as a prerequisite, and to verify that the scalings predicted by the universality class indeed capture its equilibrium behavior, or, at the very least, to evaluate the extent and nature of the departures before embarking on tests of the KZM. Of course, there is usually a microscopic theory (e.g., Bose-Hubbard), but its implication for the critical regions are typically well-established only for an infinite homogeneous system, and the extent to which it is a good approximation of an often modestly sized and inhomogeneous system   available in the laboratory is frequently not known. Moreover, it is often far from clear how to apply that theory to what is measured in the experiment (e.g., critical exponents may differ depending on how the critical region is traversed in the Mott insulator-superfluid transition\cite{Fisher89}).

To sum up, we note that the already considerable progress in verifying the KZM achieved in this millenium has accelerated in the past few years. Given the broad applicability of the KZM, it seems likely that the study of phase transition dynamics will remain an exciting research field in the foreseeable future. Our focus on experiments involving the scaling of topological defects is understandable, given the roots of the KZM. There are however other excitations of the order parameter that may be left in far-from-equilibrium state due to the KZM, and that can be used to test it. We have discussed solitons and vortices in BEC as examples, but even more transient excitations (e.g. those created in superconductors\cite{Mih1,Mih2} or left behind by the chiral symmetry breaking in $^3$He\cite{ITK13}) may be of interest in this respect.


\section*{Acknowledgments}

We thank Tom Kibble for the insightful contributions that have started the field we have attempted to review, and for many useful, pleasant, and inspiring interactions. It is also a pleasure to thank  Sang-Wook Cheong, Franco Dalfovo, Bogdan Damski,  Brian DeMarco, Piotr Deuar, Jacek Dziarmaga, Gabriele Ferrari,  Uwe Fischer, Jerome Gaunlett, Guo-Ping Guo,  Wenceslao Gonz\'alez-Vi\~nas, Paul Haljan, Valery Kiryukhin, Giacomo Lamporesi, Chuan-Feng Li, Shi-Zeng Lin, Tanja E. Mehlstaubler, Emilie Passemar, Marek M. Rams, Ray J. Rivers, Kazimierz Rzazewski, Nikolai Sinitsyn, Grigori E. Volovik, Emilia Witkowska and Vivien Zapf for useful comments and discussions.

This work is supported by the U.S Department of Energy through the LANL/LDRD Program and a  LANL J. Robert Oppenheimer fellowship (AD).

\appendix{Spontaneous Symmetry Breaking: the role of topology}\label{secSSB}


Spontaneous symmetry breaking arises in situations where a symmetry of the system is not manifested in its ground state, and it is a phenomenon tied to the degeneracy of the later \cite{Kibble76,Kibble03}. A well-known example is the breakdown of rotational invariance in a ferromagnet.
This scenario is relevant  in cosmology \cite{Kibble76,Kibble80}, elementary particle physics \cite{Coleman85,PSB03},  and condensed matter systems\cite{Kibble03,Zurek96}.
We next summarize the basics of homotopy theory and its use in this context, at a somewhat technical level.
Consider the case in which the Hamiltonian (or free-energy functional) $\hat{\mathcal{H}}$ of the system  is invariant under an operation $g$ of the symmetry group $G$, which is represented by a unitary transformation $U(g)$, 
\beqa
\label{Ginv}
U^{-1}(g)\hat{\mathcal{H}}U(g)=\hat{\mathcal{H}}, \forall g\in G.
\eeqa
Now, assume that there exists an order parameter, and operator $\hat{\psi}$ whose ground-state expectation value is not invariant under $G$, i.e.,
\beqa
\la 0|U^{-1}(g)\hat{\psi} U(g)|0\ra=D\la \hat{\psi}\ra_0\neq \la \hat{\psi}\ra_0,
\eeqa
where $D$ is a rotation matrix.
That is, the states $U(g)|0\ra$ and $|0\ra$ are nonequivalent, but are degenerate according to (\ref{Ginv}).
Typically, different phases of the system will have a symmetry group, a subgroup of $G$ called  the isotropy group $H$, which represents the leftover symmetry in the broken-symmetry phase.
An arbitrary element $h\in H$ leaves invariant the order parameter $\hat{\psi}$, $h\hat{\psi}=\hat{\psi}$.
The order parameter manifold $\mathcal{M}$ of degenerate vacuum states is homeomorphic (from the Greek for ``similar shape'', a relation denoted by the symbol $\simeq$) to the (left) coset space of $H$ in $G$ \cite{Mermin79}, 
\beqa
M\simeq G/H.
\eeqa
The simplest example is that of the linear to zigzag transition that for homogeneous ion chains is characterized by $G=\mathbb{Z}_2$, $H=e$ ($e$ being the identity $\{1\}$), and $\mathcal{M}\simeq\mathbb{Z}_2$, discussed in detail in section \ref{secKink}.  Sections \ref{secLoop}, \ref{secSoliton}, and \ref{secVortex} are devoted to the BEC transition associated with a scalar order parameter, where $G=U(1)$, $H=\{1\}$, and $G/H=U(1)$. 
Symmetry breaking in spinor Bose-Einstein condensates is more complex, and its characterization has recently led to a large body of research \cite{Uedabook,KU12,SU13}.
following the observation of spin textures in the laboratory \cite{Sadler06}. For instance, a spin-1 BEC is characterized by a $G=U(1)\times SO(3)$ 
resulting from the invariance under $U(1)$ gauge transformations and rotations in spin space, and that can lead to a variety of symmetry breaking scenarios \cite{Uedabook}.

Homotopy theory deals with continuous transformations between objects that belong to the same equivalence class and it can be used for the systematic classification of topological excitations.                                                                                                                 
Let $I=[0,1]$ and consider two continuous maps $f,g: X\rightarrow Y$ between topological spaces $X$ and $Y$. A homotopy between $f$ and $g$ is a continuous map $F:X\times I\rightarrow Y$ satisfying $F(x,0)=f(x)$, $F(x,1)=g(x)$, $\forall x\in X$. Provided $F$ exists, $f$ is said to be homotopic to $g$, which is symbolically denoted by $f\sim g$. This is an equivalence relation satisfying reflectivity ($f\sim f$), symmetry ($f\sim g$ implies $g\sim f$), and transitivity (if $f\sim g$ and $g\sim h$ then $f\sim h$). 
A path with initial point $x_0$ and final point $x_1$ is a continuous map $\alpha:I\rightarrow X$ such that $\alpha(0)=x_0$ and $\alpha(1)=x_1$. A path for which $x_0=x_1$ is called a loop with base point $x_0$, this is, a loop in which the boundary $\partial I$ of $I=[0,1]$ is mapped to $x_0$. We shall refer to two specific types of loops below, a constant loop $c:I\rightarrow X$ which has a fixed image in $X$ $\forall t\in I$, and an inverse loop $\alpha^{-1}(t)\equiv\alpha(1-t)$ $\forall t\in I$. Two loops $\alpha,\beta:I\rightarrow X$ with base point $x_0$ are homotopic ($ \alpha\sim \beta$) given that an homotopy $F:I\times I\rightarrow X$ exists,
i.e., a continuous map $F:I\times I\rightarrow X$ can be found that satisfies $F(t,0)=\alpha (t)$, $F(t,1)=\beta(t)$ $\forall t\in I$ and $F[0,t']=F[1,t']=x_0$ $\forall t'\in I$.
The set of all loops with base point $x_0$  can be classified into homotopy classes. 

A homotopy class $[\alpha]$ is the set of loops which are homotopic to $\alpha$.
The fundamental group or first homotopy group is the set of all homotopy classes of loops with base point $x_0$. It is denoted by $\pi_1(X,x_0)$ and satisfies the group properties
with respect to the product  of homotopy classes. This product is defined by  $[\alpha]\cdot[\beta]=[\alpha\cdot\beta]$, where $\alpha\cdot\beta$ is the product of loops $\alpha$ and $\beta$ in which $\alpha$ is first traversed and then $\beta$ is traversed.
Specifically, the product of homotopy classes in $\pi_1(X,x_0)$ satisfies
\beqa
& & ([\alpha]\cdot[\beta])\cdot[\gamma]=[\alpha]\cdot([\beta]\cdot[\gamma]\\
& & [\alpha]\cdot[c]=[c]\cdot[\alpha]=[\alpha]\\
& & [\alpha]\cdot[\alpha^{-1}]=[\alpha^{-1}]\cdot [\alpha]=[c]
\eeqa
where the identity element $[c]$ is given by the set of loops homotopic to a constant loop. 

There exists an isomorphism (a bijective homomorphism) between fundamental groups $\pi_1(X,x_0)$ and $\pi_1(X,x_1)$ of loops within the same connected topological spaces $X$ with different base points $x_0$ and $x_1$
which allows us to use the simplified notation $\pi_1(X)$ for the fundamental group.
A mapping from a loop to the unit circle $S^1$ is described by the isomorphism between $\pi_1(S^1)$ and $\mathbb{Z}$, 
where the integer winding number  corresponds to the number of times the loop wraps around the unit circle.
Higher homotopy groups are defined in a similar way to $\pi_1$ by considering homotopy classes of the $n$-sphere $S^n=\{x\in\mathbb{R}^{n+1}| |x|^2=1\}$. 
Let us consider the $n$-cube 
$I^n=I\times\cdots\times I =\{(s_1,\dots,s_n)|s_i\in[0,1]\,\, \forall \,0\leq i\leq n\}$ with boundary 
$\partial I^n=\{(s_1,\dots,s_n)\in I^n|s_i =0\,\, {\rm or} \,\,1\}$. 
A map $\alpha:I^n\rightarrow X$ that maps the boundary $\partial I^n$ to a point $x_0$ is a $n$-loop. 
When a homotopy exists between $n$-loops $\alpha$ and $\beta$, they are said to be homotopic, and the set of $n$-loops homotopic to a given $n$-loop $\alpha$ 
constitutes a homotopy class $[\alpha]$. The $n^{\rm th}$ homotopy group of $n$-loops with base point $x_0$ is given by the set of homotopy classes of $n$-loops .

The classification of topological excitations  is achieved by the homotopy groups $\pi_n(\mathcal{M})$ of the order parameter manifold $\mathcal{M}$ with the dimension of homotopy being given by 
$n=D-d-1$ in terms of the spatial dimension $D$ and the dimension of the (singular) topological excitation $d$ (for nonsingular topological excitations such as skyrmions, $n=D-d$).
The homotopy groups $\pi_n(\mathcal{M})$ characterize mappings from the $n$-sphere $S^n$ enclosing the topological excitation in real space into the vacuum manifold $\mathcal{M}$.   
Elements of a given group $\pi_n(\mathcal{M})$ belong to the same class of stable topological excitations, equivalent by continuous deformations. 
The number  of {\it domains} or disconnected regions in $\mathcal{M}$ is given by $\pi_0(\mathcal{M})$ (formally $\pi_0$ lacks a group structure). If $\pi_0(\mathcal{M})=k$, there are $k+1$ disconnected regions. When $\mathcal{M}$ is disconnected,  topological excitations associated with the different choices of $\la \hat{\psi}\ra_0$ in space are known as {\it domain walls}, and  are typically associated with the breakdown of a discrete symmetry, as in the linear to zigzag transition.
One can next consider  the change of the order parameter along closed loops in $S^1$ in real space within  the same connected component of $\mathcal{M}$. If $\la \hat{\psi}\ra_0$ is a smooth function along the loop, then $\pi_1(\mathcal{M})$ is trivial and equal to the identity $I$. Otherwise, elements of the group $\pi_1(\mathcal{M})\neq I$ characterize {\it line defects}  or {\it strings},  such as quantized vortices in superfluids and scalar BEC, and flux tubes in type II superconductors, associated with $U(1)$ symmetry breaking. 
Topological excitations known as {\it monopoles}  arise in the presence of non-contractive surfaces in $\mathcal{M}$ such as $S^2$, whenever $\pi_2(\mathcal{M})\neq I$. They are associated with the breakdown of nonabelian symmetries to a subgroup containing $U(1)$.
In $D=3$, $d=2$ for domains walls, $d=1$ for strings, and $d=0$ for monopoles. 
Three dimensional topological defects associated with nontrivial mappings from $S^3$ into $\mathcal{M}$ are characterized by the homotopy group $\pi_3(\mathcal{M})$ and are known as {\it textures} or non-singular solitons.

%
\begin{table}[h]
\tbl{Homotopy groups of certain vacuum manifolds.}
{\begin{tabular}{@{}cccc@{}}\toprule
$\mathcal{M}$  

                 &   $\pi_1$               &   $\pi_2$           & $\pi_3$    \\
\colrule
                 &                              &                           &                 \\
$U(1)$       &  $\mathbb{Z}$      &   $0$                  &   $0$ \\
                 &                              &                           &                 \\
$SU(n)$     &  $0$                      &   $0$                  &   $\mathbb{Z}$ \\
                 &                              &                           &                 \\
$SO(3)$     &  $\mathbb{Z}_2$  &   $0$                  &   $\mathbb{Z}$ \\
                 &                              &                           &                 \\
$S^2$       &  $0$                      &   $\mathbb{Z}$  &   $\mathbb{Z}$ \\
                 &                              &                           &                 \\
$S^3$       &  $0$                      &   $0$                  &   $\mathbb{Z}$ \\
                 &                              &                           &                 \\
$S^4$       &  $0$                      &   $0$                  &   $0$ \\
\botrule
\end{tabular}}
\label{homotopytable}
\end{table}
%
The dynamics of symmetry breaking can in principle result in hybrid configurations with a variety of topological defects with different dimensions of homotopy and which  can influence each other \cite {KLS82,Uedabook}. In that case, the classification in terms of $\pi_n$ is no longer satisfactory, but Abe homotopy groups composed of possibly noncommutative groups 
$\pi_1$ and $\pi_n$ can however account for  topological excitations with $n\geq 2$.
We refer the reader to \cite{Nakahara03} for a more detailed exposition and to \cite{Uedabook,KU12,SU13} for a thorough discussion in the context of Bose-Einstein condensates.
We close pointing out that the use of conventional homotopy groups has limitations in the classification of topological defects located on the boundary of an ordered system, for which the use of relative homotopy groups has proven to be advantageous \cite{Volovik1,Volovik2}.

\end{document}